\documentclass[amsmath,amssymb]{revtex4}
\usepackage{amssymb,amsmath,latexsym,amscd,amsfonts, bbm}
\usepackage{bm}% bold math
\usepackage{graphicx}% Include figure files

\begin{document}

%\end{document}
\title{Local random quantum circuits: ensemble CP maps and swap algebras}
%\author[PZ]{Paolo Zanardi$^\ddag,^\sharp$}
%\address{$\ddag$ Department of Physics and Astronomy, and Center for Quantum Information Science \& Technology, University of Southern California, Los Angeles, CA 90089-0484  %}
%\address{$\sharp$ Centre for Quantum Technologies,  National University of Singapore, 2 Science Drive 3, Singapore 117542}

\author{Paolo Zanardi}
\affiliation{
Department of Physics and Astronomy, and Center for Quantum Information Science \& Technology, University of Southern California, Los Angeles, CA 90089-0484}
\affiliation{    Centre for Quantum Technologies,  National University of Singapore,  2 Science Drive 3, Singapore 117542}

\begin{abstract}
We define different classes of local random quantum circuits (L-RQC) 
and  show that:
a) statistical properties of L-RQC are encoded into an associated family of completely positive maps and b)  average purity dynamics can be described by the action of these maps on operator algebras of permutations (swap algebras).  An exactly solvable one-dimensional case is analyzed to illustrate the power of the swap algebra formalism. More in general,  we prove short time area-law bounds on average purity   for  uncorrelated L-RQC and  infinite time results  for  both the uncorrelated and correlated cases. 
\end{abstract}
\maketitle

%
%A more comprehensive list (along with all the relevant definitions) of research goals will be found in the section devoted to the Proposed Research. Notice that there research goals are highlighted using boldface

\section{Introduction}
The quantum-mechanical state space of a many-body system (say  $N$ particles) is huge, as its dimension scales exponentially with $N.$
The exact specification of any  quantum state therefore requires, its face value,  an exponential number of parameters, a daunting task both numerically and
experimentally. As in classical statistical mechanics one is then naturally led to resort to {\em probabilistic methods} to explore the typical properties of {\em ensembles}
of quantum states. The question now becomes: what are the ``natural" probability distribution laws  one should consider over the state space?
A first conceptually  and mathematically appealing answer is: the group-theoretic Haar distribution \cite{rep-th}. The latter choice in fact seems to implement the principle
of minimal a priori assumptions  and at the same time allows one to exploit
%(the Occam Razor as reflected in  the Jaynes maximal entropy principle \cite{jaynes})
the powerful computational tools of representation theory \cite{rep-th}. On the other hand the Haar measure  presents problems from a quantum information theoretic  perspective.
In fact it has been known for almost a decade now that  sampling {\em exactly} the Haar distribution over a many-body Hilbert space (say  $N$ particles) by a quantum circuit made out of local gates
i.e., a quantum computer, is an exponentially (in $N$) hard task \cite{joe}. This result has been recently strengthened by considering all possible
states that are achievable by time-dependent local Hamiltonians acting for a polynomial  amount of time: the quantum states that are produced in this way 
occupy an exponentially small volume in Hilbert space \cite{illusion}. This implies that the overwhelming majority of states in Hilbert space are not physical as they can only be produced after an exponentially long time. This startling fact led the authors of \cite{illusion} to dub the Hilbert space `` a convenient illusion" (sic).

However, generic properties of Haar distributed quantum states play a crucial role
in the recent revival of interest on rigorous foundations of statistical mechanics started with  the seminal papers \cite{nino06} and \cite{andreas-nature}. 
%The questions being, as ever since the Boltzmann days, the following: {\em how can we justify the amazing effectiveness of statistical ensembles e.g., canonical, in predicting the equilibrium %expectation value of (macroscopic) observables in individual (macroscopic) physical systems?}  
%This  resurrection of some of the most long-lasting hard problems in theoretical physics has, on the one hand, an experimental motivation in a series of exciting works on the thermalization dynamics %of ensembles of ultra-cold atoms \cite{cold}. On the other hand the reasons are exquisitely theoretical being related to the emergence of new conceptual paradigm that goes under the name of {\em %typicality} \cite{nino06, andreas-nature, Reimann07}. This approach simply states that we do not {\em  need} statistical ensembles because the {\em overwhelming majority of individual quantum %states} will lead to predictions that are {\em practically indistinguishable} from those one would have obtained by postulating one or the other of the statistical ensembles
%More precisely, these authors showed that,
%thanks to the measure concentration phenomenon \cite{meas-conc} featured by the Haar measure in high dimension, {\em individual} pure quantum states lead with  overwhelming probability
%to results that are (at least locally) practically indistinguishable from the standard statistical mechanics ensembles e.g., microcanonical \cite{andreas-nature}. In this way 
%a mathematically rigorous way to justify the predictive power of these latter is achieved.
In the light of the above mentioned findings  about the unphysical nature of  Haar distributed random states,
this foundation for statistical ensembles appears to be conceptually questionable. How does one introduce a family of ``physically accessible" states endowed with a natural probability measure?
In Ref. \cite{rand-mps}  a  first attempt to undertake this endeavor was made  by introducing a family of random matrix product states, in this paper we will  focus
primarily on  {\em local random quantum circuits}  (L-QRC) families \cite{RQC1,RQC2,znidaric,viola, plenio,harrow,brandao}.
Roughly speaking  (a precise mathematical definition will be given in the next section)
a L-RQC is a random sequence of $k$ random unitary transformations (quantum gates) each of which acts on small number of localized quantum degrees of freedom. 
The states of $N$ particles obtained by the action of a L-RQC (with $k={\rm{Poly}}(N)$ ) on an initial e.g., factorized, fiducial one are now regarded as the physically accessible  ones.

%{{ The first physical motivations behind the definition of local random quantum circuits  (L-QRC) families 
%that we are going to introduce below is that they can  be used to construct ensembles  of ``physically reasonable" quantum states}}  i.e., generated by local interactions, with a natural  associated  probability measure (that is {\em not} the global Haar measure). 

A second  important physical motivation for the introduction of L-RQC comes from the perspective of quantum control and quantum information processing with limited resources.
In this case lack of ideal experimental ability may prevent one to know the precise structure of the quantum gate one has enacted as well, in the case of bounded spatial resolution,  the precise
set of local degrees of freedom e.g., qubits involved. This scenario naturally leads to a modelization in which both unitary transformations and their supports are effectively random variables.

A third compelling physical motivation behind the introduction of L-RQCs is that they  provide a natural way to ``simulate" the quantum dynamics of local random (time-dependent) Hamiltonians 
with a discrete circuit.  This, for example, can be seen by expanding the evolution generated by a local-Hamiltonian using a Suzuki-Trotter type of factorization\cite{illusion}. In this way the study of the ensembles
for circuits whose length increase unboundedly   i.e.,  the limit $k\to\infty,$ will shed light on the fundamental problem of {\em quantum equilibration} \cite{andreas-dyn}. 
Finally,  L-RQC have been recently considered in the context of  $t-$designs  theory \cite{harrow,brandao}.
Even if the latter one is a quite important area of application of L-RQCs, $t-$design theory is not the focus of the present work;
it will briefly touch upon just in the last section.
%and it will not be touched upon, but in the last brief section.

The goal of this  paper is to  provide  a mathematically precise account  of  the physical approach to  L-RQC outlined  in Refs. \cite{RQC1,RQC2}.
Generalizations of the results contained therein  and rigorous proofs of a few conjectures are presented. 

\section{Local random quantum circuits}
% \label{sec:L-RQC}}%%%%%%%%%%%%%%%%%%%%%%%%%%%%%%%%%%%%%%%%%%%%%%%%%%%%%%%%%%%%

Let us start by laying down the  mathematical framework   we are going to elaborate on. 
%The reader interested in the physical motivations and broad  background for the notion of L-RQC is referred to \cite{RQC1,RQC2,viola, plenio,harrow,brandao}
%This is a natural to formalization of  the physical approach to  local random quantum circuits (L-RQC) outlined  in Refs. \cite{RQC1,RQC2}.
Let  $\{h_i\}_{i\in\Lambda}$ be a collection of ``local" Hilbert spaces labeled by a set $\Lambda$ with finite cardinality $|\Lambda|.$
For each subset $\Omega\subset\Lambda$ we have an affiliated Hilbert space ${\cal H}_\Omega:=\otimes_{i\in\Omega} h_i,$ as well as the operator
algebra ${\mathcal A}(\Omega):={\cal B}({\cal H}_\Omega).$ If $\Omega_1\subset \Omega_2$ we will consider ${\cal A}(\Omega_1)\subset {\cal A}(\Omega_2)$
by identifying $a_1\in{\cal A}(\Omega_1)$ with $a_2=a_1\otimes\openone_{\Omega_2\setminus\Omega_1}\in{\cal A}(\Omega_2).$ In particular ${\cal A}(\Omega)\subset {\cal A}(\Lambda),\,(\forall \Omega\subset\Lambda)$
and ${\cal A}(\Lambda)\cong \bigcup_{\Omega\subset\Lambda} {\cal A}(\Omega).$
 For simplicity we will further assume that each local space has the same {\em finite} dimension $d$ i.e., $h_i\cong \mathbf{C}^d,\,(\forall i\in\Lambda$). In this case
the global state-space ${\cal H}_\Lambda$ is isomorphic to $(\mathbf{C}^d)^{\otimes\,|\Lambda|}$ 
(${\rm{dim}}\,{\cal H}_\Lambda=d^{|\Lambda|}$).
In the following we will denote by ${\rm{Tr}}_\Omega\, X,\,(X\in{\cal A}({\cal H}_\Lambda))$ the partial trace with respect ${\cal H}_\Omega,\,
(\Omega\subset\Lambda).$ Namely ${\rm{Tr}}_\Omega\colon {\cal A}({\cal H}_\Lambda)\rightarrow {\cal A}({\cal H}_{\Lambda\setminus\Omega}).$
\vskip 0.5truecm
{\bf{Definition 1}}

A {\em local random quantum circuit family} ${\mathbf C }_k[\Xi,{\cal L},q^{(k)}]$ of length $k$ is defined by the following data:
\begin{itemize}
\item[i)] A family of  probability densities $\Xi:=\{d\mu_\Omega\}_{\Omega\subset\Lambda}$ over the   groups ${\cal U}(\Omega):=\{U\in {\cal A}(\Omega)\,/\,U^\dagger=U^{-1}\}$
\item[ii)] A subset ${\cal L}$ of the power set $2^{\Lambda}.$ Elements of ${\cal L}$ will be referred to as the {\em local regions}.
\item[ii)] A probability law $q^{(k)}: {\cal L} ^k\rightarrow[0,1]\,/\, {\bf{\Omega}}:=(\Omega_1,\ldots,\Omega_k)\rightarrow q^{(k)}( {\bf{\Omega}}),\;
 \sum_{ {\bf{\Omega}}\in{\cal L}^k} q^{(k)}( {\bf{\Omega}})=1.$
%
%$$\forall \Omega_1,\ldots,\Omega_k\in{\cal L}\Rightarrow
%q^{(k)}(\Omega_1,\ldots,\Omega_k)\ge 0,\, \sum_{\Omega_1,\ldots,\Omega_k\in {\cal L}} q^{(k)}(\Omega_1,\ldots,\Omega_k)=1$$
\end{itemize}
 A local quantum circuit of length $k$ is a unitary in ${\cal U}(\Lambda)$ of the form ${{U}}_{\bf{\Omega}}:=U_k U_{k-1}\cdots U_1$ where $\forall i\in\{1,\ldots,k\}\Rightarrow$
a)$ U_i\in{\cal U}(\Omega_i)$; b)  ${\bf{\Omega}}\in{\cal L}^k.$
\vskip 0.5truecm
 Roughly speaking, the idea is that the $U_{\bf{\Omega}}$'s are unitary-valued random variables  distributed according the law
$q^{(k)}({\bf{\Omega}}) dU_{\bf{\Omega}}\, (dU_{\bf{\Omega}}=\prod_{i=1}^k d\mu_{\Omega_i}(U_{\Omega_i})).$
%\prod_{i=1}^k d\mu_\Omega(U_\Omega).$
%{\bf{Remark 1.}} 
%
Two different layers of randomness are involved in our construction of L-RQC, they are associated with  items i) and ii) in Def. 1  respectively. The first one is a stochastic process (of length $k$)
in which $k$ local regions  of the base set $\Lambda$  are selected according to the joint probability distribution $q^{(k)}.$ The second is the selection, according to the $d\mu_\Omega$'s
of $k$ unitaries acting on the state-space of each of the local regions randomly selected in the previous step.  
%{\bf{Remark 3. }}

The physical picture behind these definitions  is quite simple: the experimenter has access only to random unitaries localized over regions $\Omega$ belonging to a distinguished subset ${\cal L}\subset 2^\Lambda.$
%{\bf{Remark 2. }}
The definition of the local regions set $\cal L$ is critical and it  is where extra (physically motivated)  input is needed. The idea is that the $\Omega\in{\cal L}$ are ``small local " subsets of $\Lambda.$
In the following we will define RQCs over (hyper)graphs, in that case $\Lambda$ is the vertex set and $\cal L$ will coincide with the set of (hyper)edges; in general one may think of the local regions as sets of cardinality $O(1)$ (where $|\Lambda|$ is the large scaling quantity). Also, in the case in which $\Lambda$ is naturally equipped with a metric structure e.g., the graph theoretic  distance, the local regions
will also be required to have small diameter i.e., ${\rm{diam}}(\Omega)=O(1).$
%{\bf{Remark 3. }} 

 In the following we will almost exclusively restrict ourselves to the case where all the $d\mu_\Omega$'s are the Haar measure over ${\cal U}(\Omega).$
This symmetry assumption is crucial for our analysis  as it allows one to  resort systematically to powerful group representation theory tools \cite{rep-th}. Accordingly we will hereafter drop the $\Xi$ symbol
in the definition of ${\mathbf{C}}_k.$ Notice however that other  choices are possible. For example in \cite{plenio} the $d\mu$'s were concentrated on a finite set of {\em universal} gates.  
A first  goal for this general approach is to identify families of L-RQC that are at the same time physically motivated and amenable to rigorous mathematical analysis.
A  couple of L-RQC families that stands out as a natural target for further investigations is given  by the following types of  $q^{(k)}$
\vskip 0.5truecm
{\bf{Definition 2}}
\begin{itemize}
\item{ {\em Time dependent Markovian L- RQC}}: 
$$q^{(k)}({\bf{\Omega}})=M^{(k-1)}(\Omega_k,\Omega_{k-1})M^{(k-2)}(\Omega_{k-1},\Omega_{k-2})\cdots M^{(1)}(\Omega_2,\Omega_{1})q^{(1)}(\Omega_1)$$
 where $\{M^{(j)}\}_{j=1}^{k-1}$ are $k-1$ stochastic  $|{\cal L}|\times |{\cal L}|$ matrices and $q^{(1)}$ is a distribution over $\cal L$.
 If $M^{(j)}=M,\,(\forall j)$ we have a time-independent Markov process over ${\cal L}$.
\item{ {\em Uncorrelated L-RQC}}: $q^{(k)}({\bf{\Omega}})=\prod_{j=1}^k p^{(j)}(\Omega_j),$ 
where the $p^{(j)}$ are $k$ distribution laws over $\cal L$. If $q^{(j)}= q, \,\forall j$ we have a  time-independent uncorrelated process.
\end{itemize}
%{\bf{Remark 4.}} 

Even in these restricted setups the landscape of possibilities is huge and one has to resort, once again, to physical input to further specialize the models.
For example if $\Lambda$ is the vertex set of a graph ${\cal G}:=(\Lambda, \,E)$ and ${\cal L}=E$ one may consider $M^{(j)}$ such that $M^{(j)}(\Omega_1,\Omega_2)\neq 0$
iff the edges $\Omega_1$ and $\Omega_2$  share at least one vertex. In this way the $M^{(j)}$'s define a sort of {\em random walk} on the graph ${\cal G}.$ 
%see Sect. E).

% (the chain model). 
%{\bf{Remark 5.}} 

In passing we notice that one can use   $\mathbf{C }_k[{\cal L},q^{(k)}]$ to turn ${\cal A}(\Lambda)$ in to a {\em noncommutative probability space} \cite{nc-prob} with the expectation
  $\phi\,:\,{\cal A}(\Lambda)\rightarrow {\mathbf{C}}\,/\,A\rightarrow \phi(A),$ where
\begin{equation}
\phi(A):=  \sum_{{\bf{\Omega}}\in {\cal L}^k} q^{(k)}({\bf{\Omega}})\int dU_{\bf{\Omega}}\, \langle 
 \omega, \, U^\dagger_{\bf{\Omega}}A \,U_{\bf{\Omega}})\rangle
\label{nc-exp}
 \end{equation}
 Here $\langle A, B\rangle:={\rm{Tr}}\left (A^\dagger B\right)$ the standard Hilbert-Schmidt scalar product over ${\cal A}(\Lambda)$ 
 and $\omega$ is an distinguished (initial) density operator in ${\cal A}(\Lambda).$
 
%{\bf{ We would like to explore potential relations of L-RQC with noncommutative (free) probability and random matrix theory.}}

A physically compelling model of L-RQC that received much attention recently is based of graph structures ${\cal G}=(\Lambda,\,E)$ \cite{RQC1,plenio,harrow,brandao}. In our formalism
this amount to say that i) $\Lambda$ is the set of vertices of a graph $\cal G,$ ii) ${\cal L}=E\subset \Lambda^2$ is the set of edges of $\cal G$. {Most of  the literature on random quantum circuits so far has focused on this graph case in the time-independent uncorrelated case 2) and with $\cal G$ being the complete graph}. One-dimensional correlated cases  have been discussed in \cite{RQC1, RQC2}.
 
 \section{ Ensemble maps \label{sec:goal-funct} }
%%%%%%%%%%%%%%%%%%%%%%%%%%%%%%%%%%%%%%%%%%%%%%%%%%%%%%%%%%%
 In this section we will introduce and start to analyze a set of completely positive (CP) maps \cite{CP-maps} naturally affiliated with any L-RQC family \cite{viola,harrow}.
 These maps comprise the full statistical content of the L-RQC family and play a central role in the following of the paper.

 Using  a compact notation we can  write (\ref{nc-exp}) in the symbolic form $\phi(A)= \overline{\langle \omega,\, {\mathbf{U}}^\dagger A {\mathbf{U}}\rangle}^{{\mathbf{U}}}.$ 
  For any fixed observable $A\in{\cal A}(\Lambda)$ the mapping
$X_A\,:\,{\mathbf{U}}\rightarrow \langle \omega,\, {\mathbf{U}}^\dagger A {\mathbf{U}}\rangle$ defines a classical (commutative) random variable.
The statistics of $X_A$ describes the fluctuations of the (purely) quantum expectation $\langle \omega,\, {\mathbf{U}}^\dagger A {\mathbf{U}}\rangle$ over the ensemble of the local random circuits $\mathbf{U}$
in $\mathbf{C }_k[{\cal L},q^{(k)}].$ While the $p$-moments of the noncommutative random variables $A$ ($p\in{\mathbf{N}}$) are defined  by $\mu_p(A)=\phi(A^p)$ the moments of the corresponding classical variables are given by
%\footnote{
%Notice, however, that even this classical moments can be expressed by noncommutative (quantum) expectations. In fact
%one can write $\mu_p(X_A)=\phi_p(A^{\otimes\,p})$ where $\phi_p$ is a ``$p$-tenderized" version of (\ref{nc-exp}) over ${\cal A}(\Lambda)^{\otimes\,p}$:
 %$\phi_p(A):=\overline{ \langle \omega^{\otimes.p},\, ({\mathbf{U}}^{\otimes\,p})^\dagger A {\mathbf{U}}^{\otimes\,p}\rangle }^{\mathbf{U }}.$
%}
%\begin{equation}
$\mu_p(X_A)=\overline{X_A({\mathbf{U }})^p }^{\mathbf{U }}= \overline{ \langle \omega,\, {\mathbf{U}}^\dagger A {\mathbf{U}}\rangle^p }^{\mathbf{U }}=\langle \ \omega^{\otimes\,p},\, 
\overline{  ({\mathbf{U}}^\dagger A {\mathbf{U}})^{\otimes\,p} }^{\mathbf{U}}\rangle.
$
%\label{moments}
%\end{equation}
These moments can be conveniently  expressed in terms of a family of maps ${\cal R}_p$ associated with   $\mathbf{C }_k[{\cal L},q^{(k)}]:$ 
$\mu_p(X_A):=\langle \omega^{\otimes\,p},\, {\cal R}_p(A^{\otimes\,p})\rangle$ \cite{viola, RQC1}. Here ${\cal R}_p\,:\,{\cal A}(\Lambda)^{\otimes\,p}\rightarrow{\cal A}(\Lambda)^{\otimes\,p}\,/\, A\rightarrow \overline{  ({\mathbf{U}}^\dagger A {\mathbf{U}})^{\otimes\,p} }^{\mathbf{U}}$ are   completely positive (CP) maps \cite{CP-maps}.  More explicitly, if $A\in {\cal A}(\Lambda)^{\otimes\,p},$ one can write 
\begin{equation}
{\cal R}_p(A)=  \sum_{{\bf{\Omega}}\in {\cal L}^k} q^{(k)}({\bf{\Omega}}) {\cal R}_{p,\bf{\Omega}}(A),\quad 
{\cal R}_{p,\bf{\Omega}}(A):=\int dU_{\bf{\Omega}}  \, (U^\dagger_{\bf{\Omega}})^{\otimes\,p} A \,U_{\bf{\Omega}}^{\otimes\,p}
\label{CP}
\end{equation}
We will refer to the ${\cal R}_p$'s as the {\em ensemble maps}.
% and they play a central role in the approach advocated in this paper as {\em they comprise the full statistical content of the L-RQC family}. 
Indeed, the  distribution law for any  $X_A$'s is determined by the Fourier transform of the characteristic function $\chi_{A}(t):=\sum_{p=0}^\infty \frac{(it)^p}{p!} \mu_p(X_A).$
This can be in turn expressed as $\langle\omega_\infty, {\cal R}_\infty(t)(A_\infty)\rangle$ where $x_\infty:=\oplus_{p=0}^\infty x^{\otimes\, p}, \,(x=A,\,\omega)$
and $ {\cal R}_\infty(t):= \oplus_{p=0}^\infty \frac{(it)^p}{p!} {\cal R}_p$ is a formal CP-map over the {\em full (operator) Fock space} ${\cal A}_\infty(\Lambda)=\oplus_{p=0}^\infty {\cal A}(\Lambda)^{\otimes\,p}.$
%In the noncommutative case the expectation is given by $\phi(A)= \langle\omega, {\cal R}_1(A)\rangle,$ and therefore
%for the characteristic function of $A\in{\cal A}(\Lambda)$ one has $\chi_A(t)=\phi(e^{itA})=\langle\omega, {\cal R}_1(e^{itA})\rangle.$

The ensemble maps ${\cal R}_p$'s are clearly uniquely defined by the data in $\mathbf{C }_k[{\cal L},q^{(k)}]$
and should be characterized for  physically relevant L-RQC families. In this paper we will often consider the ${\cal R}_p$ as operators
over the Hilbert-Schmidt space ${\cal A}(\Omega)^{\otimes\,p};$ their norms are defined accordingly i.e., 
$\|{\cal R}_p\|:=\sup \{\|{\cal R}_p(A)\|/ \,\|A\|:=\sqrt{{\rm{Tr}} A^\dagger A}=1, \, A\in {\cal A}(\Omega)\}.$
Below we give
 a summary of their general properties. 
\vskip 0.5truecm
{\bf{Proposition 0}}

{\bf{i)}} The maps ${\cal R}_p$ are:  trace-preserving (${\rm{Tr}} \,{\cal R}_p(X)={\rm{Tr}}\,X$) and  unital (${\cal R}_p(\openone)=\openone$).
% and $\|{\cal R}_p\|:=\sup_{\|A\|=1} \|{\cal R}_p(A)\|=1.$

%\footnote{
{\bf{ii)}} If $q^{(k)}({\bf{\Omega}})=q^{(k)}(\tilde{\bf{\Omega}}), \,(\tilde{\bf{\Omega}}:=(\Omega_k,\ldots,\Omega_1)), \forall{\bf{\Omega}},$ the ${\cal R}_p$'s are
hermitean with respect to the Hilbert-Schmidt scalar product over ${\cal A}(\Lambda)^{\otimes\,p}.$

{\bf{iii)}} In the uncorrelated case (Def. 2) they factorize: ${\cal R}_p=\prod_{j=1}^k {\cal R}^{(j)}_p$ where $ {\cal R}^{(j)}_p(A)=\sum_{\Omega\in{\cal L}} q^{(j)}(\Omega) {\cal R}_{p,\Omega}(A),\,
 {\cal R}_{p,\Omega}(A):= \int dU_{\Omega} (U_{\Omega}^\dagger)^{\otimes\,p} A U_{\Omega}^{\otimes\,p}.$ 

{\bf{iv)}}  If  $dU_{{\bf{\Omega}}}$ is the Haar measure and  the distinguished state $\omega$ is a pure state {\em completely factorized over} $\Lambda$ 
i.e., $\omega=\otimes_{i\in\Lambda} |\phi_i\rangle\langle\phi_i|,$ then
\begin{equation}
\langle \omega_\Omega^{\otimes\,p},\,{\cal R}_{p, {\Omega}}(A)\rangle = \langle P^{(+)}_p,\, A\rangle  
\label{symm-proj}
\end{equation}
where $\omega_\Omega=\otimes_{i\in\Omega} |\phi_i\rangle\langle\phi_i|$ and $P^{(+)}_p= \frac{(d^{|\Omega|}-1)!}{(d^{|\Omega|}+p-1)!}  \sum_{\sigma\in{\cal S}_p}\sigma$ is the (normalized) projector onto the totally ${\cal S}_p$-symmetric subspace of ${\cal H}^{\otimes\,p}_\Omega$.

{\bf{v)}} In the Haar measure case the  $ {\cal R}^{(j)}_p$ are positive semi-definite operator ($\langle X, {\cal R}^{(j)}_p(X)\rangle\ge 0,\forall X$),  their spectra are contained in $[0,1]$ and always contain $1.$
%\end{itemize}

{\bf{Proof.--} } {\bf{i)}}--{\bf{iii)}}  follow easily from Eq (\ref{CP}).
 
{\bf{iv)}} In this case representation theory implies that the ${\cal R}_{p,{\Omega}}$'s (in \ref{CP}, $k=1$)
are {\em projections} on the commutant of the representations $U\in{\cal U}({\cal H}_\Omega)\rightarrow U^{\otimes\,p}\in {\cal U}({\cal H}^{\otimes\,p}_\Omega)$ \cite{rep-th}.
 By the Schur-Weyl duality this commutant
is the algebra generated by the permutations ${\cal S}_p\ni\sigma\,:\,{\cal H}^{\otimes\,p}_\Omega\rightarrow {\cal H}^{\otimes\,p}_\Omega\,/\, \otimes_{i=1}^p \phi_i\rightarrow   \otimes_{i=1}^p \phi_{\sigma(i)}$
\cite{rep-th}. Moreover,
in the computations of the moments  one has eventually to contract with  $\omega^{\otimes\,p}$ whose support is entirely contained in the  ${\cal S}_p$-symmetric subspace
of ${\cal H}^{\otimes\,p}_\Omega.$ This implies that the only relevant part of the projection
is the one associated with the identity irrep of ${\cal S}_p.$ This leads to the explicit formula (\ref{symm-proj}). 
%where $\omega_\Omega=\otimes_{i\in\Omega} |\phi_i\rangle\langle\phi_i|$ and $P^{(+)}_p= \frac{(d^{|\Omega|}-1)!}{(d^{|\Omega|}+p-1)!}  \sum_{\sigma\in{\cal S}_p}\sigma$ is the (normalized) projector onto the %totally ${\cal S}_p$-symmetric subspace of ${\cal H}^{\otimes\,p}_\Omega$.
%A special case of (\ref{symm-proj}) plays a central role in the results on typical entanglement in physical states of Refs. \cite{RQC1,RQC2} (see Sect. C). A key technical question is then 
%{\bf{How  Eq. (\ref{symm-proj}) extends   to  non factorized and mixed initial states? 
 %}}

{\bf{v)}}  In the Haar measure case each of the $ {\cal R}^{(j)}_p$ in iii) of Prop. 0
is a convex combination of the projectors ${\cal R}_{p,\Omega}$ and it is therefore a  positive semi-definite.
Moreover , $\|{\cal R}^{(j)}_p\|=\|\sum_{\Omega\in{\cal L}} q^{(j)}(\Omega) {\cal R}_{p,\Omega}\|\le \sum_{\Omega\in{\cal L}} q^{(j)}(\Omega) \|{\cal R}_{p,\Omega}\|=
 \sum_{\Omega\in{\cal L}} q^{(j)}(\Omega) =1,$
hence Sp$({\cal R}^{(j)}_p)\subset [0,\,1]. $ Since e.g,  $\openone$ is always a fixed point $1\in{\rm{Sp}}({\cal R}^{(j)}_p).$ $\hfill\Box$
%Their spectra are then in $[0,1]$ and always contain $1$ (e.g., $\openone$ is a fixed point).  $\hfill\Box$

\section{First moments}%%%%%%%%%%%%%%%%%%%%%%%%%%%%%%%%%%%%%%%%%
In this section we will analyze the L-RQC first moments of quantum observables by means of  the ensemble maps ${\cal R}_1.$ In particular the dynamics of these first moments can be studied in the limit
in which the circuit size grows unboundedly i.e., $k\to\infty.$

Let $\omega\in{\cal A}(\Lambda)$ be a density matrix (i.e., $\omega\ge0,\, {\rm{Tr}(}\omega)=1$) and $A\in{\cal A}(\Lambda)$ a quantum observable. 
%In this section we will study the action of the maps ${\cal R}_1$ for the uncorrelated L-RQC case.  
The expectation value of the random variable ${\mathbf{U}}\mapsto\langle\omega,\,{\mathbf{U}}^\dagger A{\mathbf{U}}\rangle$
over the ensemble of {\em uncorrelated, time-independent} circuits of length $k$ is given by (see previous Sect. ) by $\phi_k(A):=\langle\omega,\,{\cal R}_1^k(A)\rangle,$ where ${\cal R}_1=\sum_{\Omega\in{\cal L}} q^{(1)}(\Omega) {\cal R}_{1,\Omega}.$ We will also assume that ${\cal L}$ is a covering of $\Lambda$ i.e., $\cup_{\Omega\in{\cal L}}\Omega=\Lambda.$
\vskip 0.5truecm
{\bf{Proposition 1}}

{\bf{i)}} In the Haar measure case the ensemble maps ${\cal R}_{1,\Omega}$ are projections:
${\cal R}_{1,\Omega}(X)=d^{-|\Omega|} \openone_\Omega\otimes {\rm{Tr}}_\Omega(X).$

{\bf{ii)}} $ {\cal R}_{1,\Omega_1}{\cal R}_{1,\Omega_2}={\cal R}_{1,\Omega_2}{\cal R}_{1,\Omega_1}= {\cal R}_{1,{\Omega_1\cup\Omega_2}}.$

{\bf{iii)}}  ${\rm{Sp}}({\cal R}_1)\subset\{  q(   {\underline\alpha}):= \sum_{k=}^{|{\cal L}|} q^{(1)}(\Omega_k)\alpha_k\,/\,
\alpha_1,\ldots,\alpha_{|{\cal L}|}\in{\mathbf{Z}}_2=\{0,\,1\}\}\ni 1$

{\bf{iv)}} 
The following  estimate for  the convergence $\phi_k(A)\rightarrow\phi_\infty(A)= d^{-|\Lambda|} {\rm{Tr}} A,$ holds
\begin{equation}
k\gg\frac{  \log(\|\omega\|_2 \|A\|_2/\epsilon) + (|{\cal L}|-1)\log 2}{\log\frac{1}{1-  q^{(1)}(\Omega_*) }}\Rightarrow |\phi_k(A)-\phi_\infty(A)|\le \epsilon
\label{crude}
\end{equation}
where $\Omega_*=\arg\min_\Omega q^{(1)}(\Omega).$

{\bf{Proof.--}}

{\bf{i)}}  It follows from the irreducibility of the representation of the unitary group of ${\cal A}(\Omega)$ over ${\cal H}_\Omega$
\cite{rep-th}.  In particular if $X=X_\Omega\otimes X_{\Omega^c}\in{\cal A}(\Omega)\otimes{\cal A}(\Omega^c)$ one has ${\cal R}_{1,\Omega}(X)=d^{-|\Omega|} \openone_\Omega\otimes  X_{\Omega^c}\,{\rm{Tr}}(X_\Omega).$ 
 
{\bf{ii)}} It is a direct computation, if $X\in{\cal A}(\Omega)$ then ${\cal R}_{1,\Omega_1}{\cal R}_{1,\Omega_2}(X)$ is given by
\begin{eqnarray*} 
 d^{-|\Omega_2|} R_{1,\Omega_1}(\openone_{\Omega_2} \otimes {\rm{Tr}}_{\Omega_2}(X)) &=& 
d^{-(|\Omega_1| + |\Omega_2|)} \openone_{\Omega_1} \otimes {\rm{Tr}}_{\Omega_1}(\openone_{\Omega_2} \otimes {\rm{Tr}}_{\Omega_2}(X)) \\ &=& d^{-(|\Omega_1| + |\Omega_2|)} \openone_{\Omega_1} \otimes d^{|\Omega_1 \cap \Omega_2|} \openone_{\Omega_2 \setminus \Omega_1} {\rm{Tr}}_{\Omega_1 \cup \Omega_2}(X) \\ &=& d^{-|\Omega_1 \cup \Omega_2|} \openone_{\Omega_1 \cup \Omega_2} \otimes {\rm{Tr}}_{\Omega_1 \cup \Omega_2}(X). \end{eqnarray*}
Where we have used: $\openone_{\Omega_2}=\openone_{\Omega_2\setminus\Omega_1 }\otimes \openone_{\Omega_2\cap\Omega_1 },\,
\openone_{\Omega_1\cup\Omega_2}=\openone_{\Omega_1 }\otimes \openone_{\Omega_2\setminus\Omega_1 },$
${\rm{Tr}}\,  \openone_{\Omega_2\cap\Omega_1 }=d^{|\Omega_2\cap\Omega_1|},$
and $|\Omega_1 \cup \Omega_2|=|\Omega_1|+|\Omega_2|-|\Omega_2\cap\Omega_1|.$ Of course an identical result is obtained
for ${\cal R}_{1,\Omega_2}{\cal R}_{1,\Omega_1}(X).$ 

%{\bf{ii)}}  When $\Omega_1\cap\Omega_2=\emptyset$ the claim is obvious.
%Let us assume then $\Omega_1\cap\Omega_2\neq\emptyset$ and  w.l.o.g.  that $\Lambda=\Omega_1\cup\Omega_2.$
%We write $\Lambda=\Lambda_1\cup \Lambda_2\cup \Lambda_3$ where $\Lambda_1:=\Omega_1\setminus\Omega_2,\,%\Lambda_2=\Omega_1\cap\Omega_2,$ and $\Lambda_3:=\Omega_2\setminus\Omega_1.$
%Let us consider $X\in{\cal A}(\Lambda)=\otimes_{i=1}^3 {\cal A}(\Lambda_i)$ of the form $X_1\otimes X_2\otimes X_3.$
%One has ${\cal R}_{1,\Omega_1}(X)= d^{-|\Omega_1|} \openone_{\Lambda_1}\otimes \openone_{\Lambda_2}\otimes X_3\, {\rm{Tr}} X_1 {\rm{Tr}} %X_2,$ therefore
%${\cal R}_{1,\Omega_2} ({\cal R}_{1,\Omega_1}(X))= d^{-|\Omega_2|} d^{-|\Omega_1|} \openone_{\Lambda_1}\otimes %\openone_{\Lambda_2}\otimes \openone_{\Lambda_3}
%{\rm{Tr}} \openone_{\Lambda_2}\, {\rm{Tr}}X_3\, {\rm{Tr}} X_1 \,{\rm{Tr}} X_2,=   d^{-|\Omega_1|-|\Omega_2|+|\Omega_1\cap\Omega_2|}\, %\openone_{\Lambda}\, \prod_{i=1}^3{\rm{Tr}} X_i=
%d^{-|\Lambda|} \openone_\Lambda {\rm{Tr}} X,\,(|\Lambda|=|\Omega_1\cup\Omega_2|= |\Omega_1|+|\Omega_2|-|\Omega_1\cap\Omega_2|.%$
%Computing ${\cal R}_{1,\Omega_1} ({\cal R}_{1,\Omega_2}(X))$ clearly gives the same result. 

{\bf{iii)}}  Since ${\cal R}_1$ is a convex combination of commuting projections its diagonalization is formally immediate. Let us introduce the notations
$P_k^{(0)}:=1-{\cal R}_{\Omega_k},$ and $P_k^{(1)}:={\cal R}_{\Omega_k},\,(\{\Omega_1,\ldots,\Omega_{|{\cal L}|}\}={\cal L}).$
One can then decompose the identity CP map as follows $1=\prod_{k=1}^{|{\cal L}|} (\sum_{\alpha_k=0,1} P^{(\alpha_k)}_k)=
\sum_{\alpha_1,\ldots,\alpha_{|{\cal L}|}=0,1} \prod_{k=1}^{|{\cal L}|}  P^{(\alpha_k)}_k.$  Therefore
\begin{equation}
{\cal R}_1={\cal R}_1\cdot 1=\sum_{\alpha_1,\ldots,\alpha_{|{\cal L}|}=0,1}  \left(   \sum_{k=1}^{|{\cal L}|} q^{(1)}(\Omega_k)\alpha_k  \right) \prod_{k=1}^{|{\cal L}|}  P^{(\alpha_k)}_k
=: \sum_{ {\underline\alpha}\in{{\mathbf{Z}}_2^{|{\cal L}|}} } q(   {\underline\alpha}) P[{\underline\alpha}]
\label{R-decomp}
\end{equation}
where we used ${\cal R}_{1,\Omega_k} P[\underline{\alpha}]=\alpha_k P[\underline{\alpha}] (\forall\underline{\alpha}\in{\mathbf{Z}}_2^{|{\cal L}|}).$
Notice that not all the projectors $ P[{\underline\alpha}]:=\prod_{k=1}^{|{\cal L}|}  P^{(\alpha_k)}_k$ are necessarily non zero
\footnote{For example: $\Omega_1\subset\Omega_2\Rightarrow {\rm{Im}}\,{\cal R}_{\Omega_1}\supset   {\rm{Im}}\,{\cal R}_{\Omega_2}\Rightarrow {\cal R}_{\Omega_2}(1-{\cal R}_{\Omega_1})=0.$
}, in any case 
${\underline{\alpha}}:=(\alpha_1,\ldots,\alpha_{\alpha_{|{\cal L}|}})\neq {\underline{\alpha}}^\prime\Rightarrow P [{\underline\alpha}] P[{\underline\alpha}^\prime]=0.$
This show that the decomposition (\ref{R-decomp}) is a spectral one from which the claim  Sp$({\cal R}_1)\subset \{q(\underline\alpha)\}_{\underline{\alpha}\in{\mathbf{Z}}_2^{|{\cal L}|}}$ follows.
% of ${\cal R}_1$ is contained in the set $\{  q(   {\underline\alpha}):= \sum_{k=}^{|{\cal L}|} q^{(1)}(\Omega_k)\alpha_k\,/\,
%\alpha_1,\ldots,\alpha_{|{\cal L}|}\in{\mathbf{Z}}_2=\{0,\,1\}\}.$

 Since $\exists X$ such that $P[(1,\ldots,1)](X)=\prod_{k=1}^{|{\cal L}|} {\cal R}_{1,\Omega_k}(X)= {\cal R}_{1,\cup_{k=1}^{|{\cal L}|}\Omega_k}(X)={\cal R}_{1,\Lambda}(X)=d^{-|\Lambda|} \openone_\Lambda\, {\rm{Tr}}X\neq 0$ one sees that   $P[(1,\ldots,1)]\neq 0$ and therefore
 $q(1,\ldots,1)=\sum_{\Omega}  q^{(1)}(\Omega)=    1$ always belong to the spectrum of ${\cal R}_1.$
Moreover (Prop. 0, v) 1 is the maximum eigenvalue
and $q(\alpha)=1-\sum_{i:\alpha_i=0} q^{(1)}(\Omega_i)<1$ if $\alpha\neq (1,\ldots,1).$

{\bf{iv)}}
We first observe  that the second largest $q(\alpha)$ is upper bounded by 
%$\lambda_2={\rm{max}}\{ q( {\underline\alpha})\,/\, \underline\alpha\neq(1,\ldots,1) {\rm{and}} P[\underline\alpha]\neq 0\}$ and therefore
%it is bounded above by 
$1-q^{(1)}(\Omega_*)$ where $\Omega_*=\arg\min_\Omega q^{(1)}(\Omega).$ 
Second, ${\cal R}_1^k(A) =\sum_{\underline{\alpha}} q(\underline{\alpha})^k P[\underline{\alpha}](A)=P[(1\ldots,1)](A) +\sum_{\underline{\alpha} \neq(1,\ldots,1)} q(\alpha)^k P[\underline{\alpha}](A)$
from which $\lim_{k\to\infty} {\cal R}_1^k=P[1,\ldots,1]$ and  
\begin{eqnarray}
 |\phi_k(A)-\phi_\infty(A)|&=&|\phi_k(A)-\langle\omega,\,P[1,\ldots,1](A)\rangle|\le \sum_{\underline{\alpha} \neq(1,\ldots,1)} q(\alpha)^k | \langle \omega, P[\underline{\alpha}](A)\rangle|
\nonumber \\
&\le& (1-q^{(1)}(\Omega_*))^k  \sum_{\underline{\alpha} \neq(1,\ldots,1)}
\|P[\underline{\alpha}]\|\|\omega\|_2 \|A\|_2 \le  (1-q^{(1)}(\Omega_*))^k  2^{|{\cal L}|-1} \|\omega\|_2 \|A\|_2.\le\epsilon
\label{ineq}
\end{eqnarray}
Here we  used Cauchy-Schwarz $|\langle\omega, P[{\underline{\alpha}} ](A)\rangle|\le \|\omega\|_2 \| P[{\underline{\alpha}}](A)\|_2\le  \|\omega\|_2 \|P[{\underline{\alpha}}]  \|A\|_2\le \|\omega\|_2 \|A\|_2$ as $P[{\underline{\alpha}}]$
is a projector i.e., norm one . Solving the last inequality in (\ref{ineq}) for $k$ gives the estimate
(\ref{crude}). $\hfill\Box$
\vskip 0.5truecm
Whereas the asymptotic value $\phi_\infty(A)$ does not depend on the distribution $q^{(1)}$ (as long as $\cup_{\Omega\in{\cal L}}\Omega=\Lambda)$
the convergence rate may well do so. In order have a faster convergence one would like to make $q(\Omega_*)$ as large as possible i.e., $q(\Omega_*)=1/|{\cal L}|.$
In this case a weaker form of (\ref{crude}) gives $k\gg1/q(\Omega_*)\log(\|\omega\|_2 \|A\|_22^{|{\cal L}|-1}/\epsilon)=O( |{\cal L}|\log(\|\omega\|_2 \|A\|_2/\epsilon) +|{\cal L}|^2\log2 )\rightarrow 
|\phi_k(A)-\phi_\infty(A)|\le\epsilon.$
In several relevant examples one has $|{\cal L}|=O(|\Lambda|)$ so that (\ref{crude}) gives, for $\|A\|_2=O(1),$ a convergence estimate {\em polynomial} in the system size.

Of course (\ref{crude}) is a {\em very} crude estimate that may be strongly improved yielding  faster convergence rates. For example if $A\in{\cal A}(\Omega)$ 
then $P[\underline{\alpha}] A=0$ for all the $\underline{\alpha}'$s such that $\alpha_{k}=0$ when $\Omega_k\subset \Omega^c.$ 
This constraint greatly reduces the number of non vanishing terms in (\ref{ineq}) giving rising to a better bound. In fact,  
If  $|\{\Omega^\prime\in{\cal L}\,/\, \Omega^\prime\subset\Omega^c\}|=O(|\Omega^c|)$
then $|{\cal L}|$ in (\ref{crude}) becomes $|{\cal L}|-O(|\Omega^c|)=O(|\Omega|).$

 \section{Purity dynamics and swap algebras  \label{sec:purity-dyn}}
%%%%%%%%%%%%%%%%%%%%%%%%%%%%%%%%%%%%%%%%%%%%%%%%%%%%%%%%%%%
In this section we will show how the ensemble maps ${\cal R}_2$ allow one to study the dynamics of average purity and how this can be done in terms of
dynamically closed subalgebras of permutations. Uncorrelated as well as correlated L-RQC will be considered.

 Given a density matrix $\omega\in{\cal A}(\Lambda)$  the reduced state $\omega_\Omega\in{\cal A}(\Omega)$ associated with the region $\Omega\subset\Lambda$ is given by the partial trace $\omega_\Omega={\rm{Tr}}_{\Omega^c}(\omega)\, (\Omega^c:=\Lambda\setminus\Omega).$ A family of non-linear functions that plays an important role in quantum information theory is given by the $\alpha$-Renyi  entropies: $S_\alpha(\omega_\Omega):= -\frac{1}{\alpha-1}\log {\rm{Tr}} (\omega^\alpha_\Omega),\,(\alpha\in{\mathbf{R}}^+_0).$ In particular $\lim_{\alpha\to 1} S_\alpha(\omega_\Omega)=- {\rm{Tr}}
\left(\omega_\Omega\log\omega_\Omega\right)$ is the von Neumann entropy that for pure $\omega$ quantifies {\em quantum entanglement} between the two regions $\Omega$ and $\Omega^c$ 
\cite{qip}. 
The following, well-known, Lemma will be used  many times in this work.

{\bf{Lemma}}

For $\alpha=p\in{\mathbf{N}}$ one can write ${\rm{Tr}} (\omega^p_\Omega)={\rm{Tr}}\left(T^{(p)}_\Omega \omega^{\otimes\,p}      \right)$
where $T^{(p)}_\Omega\,:\,  {\cal H}^{\otimes\,p}_\Lambda\rightarrow{\cal H}^{\otimes\,p}_\Lambda$
%\,/\, |\phi_\Omega\rangle^{\otimes\,p}\otimes|\phi_{\Lambda\setminus\Omega}\rangle^{\otimes\,p}\rightarrow
%\left( \otimes_{i=1}^p      \right)$ 
is the cyclic permutation $\pi\,:\,i\rightarrow i+1\,{\rm{mod}}\, p$ acting on the ${\cal H}_\Omega$ factors of ${\cal H}^{\otimes\,p}_\Lambda\cong{\cal H}^{\otimes\,p}_\Omega\otimes{\cal H}^{\otimes\,p}_{\Lambda\setminus\Omega}.$  

{\bf{Proof.--}}
 Set $X:=\omega_\Omega,$ one has
\begin{eqnarray}
{\rm{Tr}} (X^p)&=& \sum_{i_1,\ldots,i_p} X_{i_1,i_2} X_{i_2,i_3}\cdots X_{i_p, i_1}= \sum_{i_1,\ldots,i_p} {\rm{Tr}}\left( |i_2,\,i_3,\ldots, i_1 \rangle\langle i_1, i_2,\ldots,i_p | X^{\otimes\,p}  \right)\nonumber\\
&=& {\rm{Tr}}\ ( \sum_{i_1,\ldots,i_p}  |i_2,\,i_3,\ldots, i_1 \rangle\langle i_1, i_2,\ldots,i_p | X^{\otimes\,p}  ) ={\rm{Tr}}\left(T^{(p)} X^{\otimes\,p}\right).
\end{eqnarray}
The result the follows by trivial extension of $T^{(p)}$ to any auxiliary tensor factor as ${\cal H}^{\otimes\,p}_{\Lambda\setminus\Omega}.$

$\hfill\Box$
\vskip 0.5truecm

We will call $T^{(2)}_\Omega$  the {\em swap} associated with $\Omega$ and it will be denoted  by $T_\Omega,$ we will also
write ${\cal R}_{2, \bf{\Omega}}$ simply as ${\cal R}_{\bf{\Omega}}$ and so on.
 Since ${\cal H}_\Omega=\otimes_{i\in\Omega} h_i$ one has
$T_\Omega= \otimes_{i\in\Omega} T_i$ where the $T_i$ act as $\pi$ in $h_i^{\otimes\,2}.$ 
The case $p=2$  has  special relevance as the 2-Renyi entropy $S_2(\omega_\Omega)=-\log {\rm{Tr}}(\omega_\Omega^2)$
is a convenient way to quantify quantum entanglement that is receiving a constantly growing attention from the quantum information and theoretical condensed matter communities see e.g., \cite{Renyi}.
If the initial state $\omega$ is evolved by a RQC $\mathbf{U}$ in $\mathbf{C }_k[{\cal L},q^{(k)}]$ the corresponding {\em purity} 
%${\rm{Tr}}\left({\mathbf{U}}^\dagger T_\Omega{\mathbf{U}}\,  \omega^{\otimes\,2})$ 
becomes a random variable whose expectation value $P_k:=\overline{  {\rm{Tr}}\left(({\mathbf{U}}^{\otimes\,2})^\dagger T_\Omega{\mathbf{U}}^{\otimes\,2}\,  \omega^{\otimes\,2}\right) }^{\mathbf U}$ is given by: 
\begin{equation}
P_k:= \sum_{{\bf{\Omega}}\in {\cal L}^k} q^{(k)}({\bf{\Omega}}) \langle\omega^{\otimes\,2},\,{\cal R}_{\bf{\Omega}}(T_\Omega)\rangle
=\langle\omega^{\otimes\,2},\, {\cal R}(T_\Omega)\rangle
\label{pur-av}
\end{equation}
By convexity it follows that $\overline{S_2}^{\mathbf U}\ge -\log P_k$, namely the negative of the logarithm of (\ref{pur-av}) provides a lower bound to the average
entanglement generated by evolving $\omega$ with RQC in $\mathbf{U}$ in $\mathbf{C }_k[{\cal L},q^{(k)}].$
Two of the key technical insights in \cite{RQC1} on which we would like to build upon in this paper are contained in the next two propositions: 
\vskip 0.5truecm
{\bf{ Proposition 2}}

The swaps $T_{\Omega}$ form an abelian group ${\cal T}_\Lambda$ of order $2^{|\Lambda|}$ whose elements have degree two i.e., $T_{\Omega}^2=\openone.$
\begin{equation}
T_{{{\Omega}}_1} T_{{{\Omega}}_2}=T_{ \Omega_1\Delta \Omega_2  },\,  
\Omega_1\Delta \Omega_2:=(\Omega_1\setminus \Omega_2)\cup  (\Omega_2\setminus\Omega_1)\,(\forall\Omega_1,\Omega_2\subset \Lambda)
\label{SA-law}
\end{equation}
The group algebra $\mathbf{C}{\cal T}_{\Lambda}$ is  $2^{|\Lambda|}$-dimensional abelian subalgebra of $\mathbf{C}{\cal T}_{\Lambda}\subset {\cal A}(\Lambda)^{\otimes\,2}$ 
under multiplication.  $\mathbf{C}{\cal T}_{\Lambda}$ will referred to as the {\em swap algebra of $\Lambda$}.

{\bf{Proof.--}} $T_{\Omega_1} T_{\Omega_2}=\left(\prod_{i\in\Omega_1}T_i\right)\left(\prod_{j\in\Omega_2}T_j\right)=\left(\prod_{i\in\Omega_1\setminus\Omega_2}T_i\right)
\left(\prod_{i\in\Omega_1\cap\Omega_2}T^2_i\right)\left(\prod_{i\in\Omega_2\setminus\Omega_1}
T_i\right)=\left(\prod_{i\in(\Omega_1\setminus\Omega_2)\cup(\Omega_2\setminus\Omega_1)}T_i\right):= T_{\Omega_1\Delta\Omega_2}.$
Where we have used $T_i^2=\openone,\, \forall i\in\Lambda.$ Notice also that the $T_\Omega$'s are linearly independent in ${\cal A}(\Lambda)^{\otimes\,2}$
as each of them amounts to a permutation of the product state basis of ${\cal H}_{\Lambda}^{\otimes\,2}\cong \otimes_{i\in\Lambda}h_i^{\otimes\,2}.$
$\hfill\Box$
\vskip 0.5truecm
Eq. (\ref{SA-law}) shows that  ${\cal T}_\Lambda$   is isomorphic to the power set of $\Lambda$ endowed with the internal operation of symmetric set difference $\Delta.$  The isomorphism being given by $\Omega\rightarrow T_\Omega.$ Moreover, the swap algebra  $\mathbf{C}{\cal T}_{\Lambda}$  is isomorphic, as a vector space, to the
space $({\mathbf{C}}^2)^{\otimes\,|\Lambda|}$ associated with 
$|\Lambda|$ qubits ($T_\Omega\cong \otimes_{j\in\Lambda} |\chi_\Omega(j)\rangle,$ $\chi_\Omega$ characteristic function of $\Omega\subset\Lambda$ ).
 Next proposition shows that  $\mathbf{C}{\cal T}_{\Lambda}$ is {\em invariant} under  the ${\cal R}_{{{\Omega}}}$'s. 
% (as opposed to the $(d^2)^{|\Lambda|}$ original space).
\vskip 0.5truecm
{\bf{Proposition 3}}%%%%%%%%%%%%%%%%%%%%%%%%%%%%%%%%%%%%% PROPOSITION 3

{\bf{i)}}  If $\Omega\subset\Lambda$ let us  define its boundary (in $\cal L$) as $\partial\Omega:=\{\Omega^\prime\in {\cal L}\,/\, \Omega^\prime\cap\Omega\neq\emptyset \,\wedge \,\Omega^\prime\cap\Omega^c\neq\emptyset\},$ then 

a)$\Omega_1\notin\partial\Omega\Rightarrow $  ${\cal R}_{\Omega_1}(T_{\Omega})=T_{\Omega}$,  otherwise 
b)
\begin{equation}
{\cal R}_{\Omega_1}(T_{\Omega})=\alpha_+T_{\Omega\setminus\Omega_1}+\alpha_- T_{\Omega\cup\Omega_1},
\label{SA-inv}
\end{equation}
where $\alpha_{\pm}=(c^+\pm c^-)/2,\, c^{\pm}=(d^{A}\pm d^{B})/(d^{A+B}\pm 1),\,
A=|\Omega_1\setminus\Omega|,\,B= |\Omega\cap\Omega_1|.$
%\end{itemize}

{\bf{ii)}} The  swap algebra is invariant under all the ensemble maps ${\cal R}.$ 

{\bf{iii)}} Let $P\in L(\mathbf{C}{\cal T}_{\Lambda})$ s.t. $T_\Omega\mapsto T_{\Omega^c},\,(\Omega\subset\Lambda)$ then $[{\cal R}_{\Omega_1},\,P]=0,\,(\forall\Omega_1\subset\Lambda)$

{\bf{Proof.--}} {\bf{i)}}
a) If $\Omega_1\subset \Omega^c$ then the result follows trivially from ${\cal R}_{\Omega_1}(\openone_{\Omega_1})=\openone_{\Omega_1}.$ If $\Omega_1\subset \Omega$
then it follows from $U_{\Omega_1}^{\dagger\otimes\,2} T_\Omega U_{\Omega_1}^{\otimes\,2}= T_{\Omega\setminus\Omega_1} U_{\Omega_1}^{\dagger\otimes\,2} T_{\Omega_1} U_{\Omega_1}^{\otimes\,2}=
T_{\Omega\setminus\Omega_1}T_{\Omega_1} \left(T_{\Omega_1}U_{\Omega_1}^{\dagger\otimes\,2} T_{\Omega_1} \right)U_{\Omega_1}^{\otimes\,2}=T_\Omega U_{\Omega_1}^{\dagger\otimes\,2}
 U_{\Omega_1}^{\otimes\,2}=T_\Omega.$

b)  First, one has  that  ${\cal R}_{\Omega_1}(T_\Omega)= T_{\Omega\setminus\Omega_1} {\cal R}_{\Omega_1}(T_{\Omega\cap\Omega_1}).$ In order to compute ${\cal R}_{\Omega_1}(T_{\Omega\cap\Omega_1}),$
notice that ${\cal R}_{\Omega_1}$ (restricted on ${\cal H}_{\Omega_1}^{\otimes\,2}$) is a projection on the  2-dimensional algebra spanned by $\openone$ and $T_{\Omega_1}$ \cite{rep-th}.
A Hilbert-Schmidt orthonormal basis of this space is given by the elements
$$
F_{\pm}:=\frac{ \openone \pm T_{\Omega_1}}{\sqrt{2d^{|\Omega_1|}( d^{|\Omega_1|}\pm 1)  }}.
$$
Hence ${\cal R}_{\Omega_1}(T_{\Omega\cap\Omega_1})=
\sum_{\alpha=\pm} F_\alpha {\rm{Tr}}\left(F_\alpha T_{\Omega\cap\Omega_1}\right).$
Using (\ref{SA-law}) one obtains, $ {\rm{Tr}}\left(F_\alpha T_{\Omega\cap\Omega_1}\right)\sim{\rm{Tr}}\left( T_{\Omega\cap\Omega_1}+\alpha T_{\Omega1-\Omega}\right)=
d^{|\Omega\cap\Omega_1|} d^{2|\Omega_1\setminus\Omega|} +\alpha d^{|\Omega_1\setminus\Omega|}  d^{2|\Omega\cap\Omega_1|}.$
Now the claim  follows by noticing that $|\Omega_1|=|\Omega_1\setminus\Omega| +|\Omega\cap\Omega_1|$ and that, from Eq. (\ref{SA-law}), one has $ T_{\Omega\setminus\Omega_1} T_{\Omega_1}=T_{\Omega\cup\Omega_1}.$

{\bf{ii)}} It is immediate by observing that  ${\cal R}_{{\bf{\Omega}}}={\cal R}_{\Omega_1}\circ {\cal R}_{\Omega_2}\circ\cdots
\circ {\cal R}_{\Omega_k}$ (remind that ${\bf{\Omega}}=(\Omega_1,\ldots,\Omega_k)\in{\cal L}^k$) and using (\ref{CP}) and (\ref{SA-inv}).

{\bf{iii)}} First notice that (from i)) $\alpha^\pm(\Omega,\Omega_1)=\alpha^\mp(\Omega^c,\Omega_1),$ whence
${\cal R}_{\Omega_1}( PT_\Omega)={\cal R}_{\Omega_1}( T_{\Omega^c})=\alpha^+(\Omega^c,\Omega_1) T_{\Omega^c\cap\Omega_1^c} +\alpha^-(\Omega^c,\Omega_1) T_{\Omega^c\cup\Omega_1}=\alpha^-(\Omega,\Omega_1)T_{(\Omega\cup\Omega_1)^c}+\alpha^+(\Omega,\Omega_1)T_{(\Omega\cap\Omega_1^c)^c}=
\alpha^-(\Omega,\Omega_1)PT_{\Omega\cup\Omega_1}+\alpha^+(\Omega,\Omega_1)PT_{\Omega\cap\Omega_1^c}=P{\cal R}_{\Omega_1}(T_\Omega).$
$\hfill\Box$
\vskip 0.5truecm

Proposition 3 is an important result. It shows that (average) purity dynamics can be mapped onto a dynamical problem  in a space comprising just $|\Lambda|$ qubits.
This mapping entails a dimensional reduction $(d^2)^{|\Lambda|}\mapsto 2^{|\Lambda|}.$ Exploiting symmetries this reduction can be made, in some cases, 
 even stronger. To start with, map $P$ in iii) is an involution i.e., $P^2=\openone,$ and therefore $\mathbf{C}{\cal T}_{\Lambda}$ breaks up in two orthogonal $(|\Lambda| -1)$-qubits subspaces (corresponding to eigenvalues $\pm 1$ of $P$) that are invariant under all ensemble maps. 
%[Notice that in spin $1/2$ language $P=\prod_{i\in \Lambda} \sigma^x_i.$].  In the entanglement dynamics with totally  factorized input one can restrict to the $+1$ eigenspace. 
Moreover, the action of the ensemble maps on the swap algebra can be highly reducible
thus lowering even further  the dimensionality of  the purity dynamics problem. 
 References \cite{RQC1} and \cite{RQC2} contain severale examples (see two below) in which the entanglement dynamics is approached
and solved at the swap algebra level because of the dramatic decrease of computational complexity.

%{\bf{Example 1.}} 
%{\bf{a)}} Let ${\cal G}=(\Lambda, E)$ be a one-dimensional chain of length $L:$
%$\Lambda:=\{ 1,\dots,\, L\},\, E=\{ \{1,2\},\,\{2,\,3\},\ldots,\{L-1,\,L\}\}.$ From Eq (\ref{SA-inv}) 
%it follows that the $(L+1)$-dimensional subspace of $\mathbf{C}{\cal T}_\Lambda$ spanned by $\{T_\Omega\,/\, \Omega=\{1,\ldots, i\},\,i=1,\ldots L\}$
%and $T_\emptyset=\openone$  is invariant under all the edge maps ${\cal R}_e\,(e\in E).$
%{\bf{b)}} Let ${\cal G}$ the {\em complete graph} with $L$ vertices. In \cite{RQC2} it has been shown that purity dynamics can be restricted to the $(L+1)$- dimensional ${\cal S}_L$-symmetric subspace of   
%$\mathbf{C}{\cal T}_\Lambda\cong (\mathbf{C}^2)^{\otimes\,L}.$ 
%In these two cases one has exponential dimensional reductions $O(d^{2L})\rightarrow O(L).$ 

%\end{equation}

%Then  Eq. (\ref{SA-inv})  simplifies to ${\cal R}_{\Omega_1}(T_{\Omega})=N_d(T_{\Omega\setminus\Omega_1}+ T_{\Omega\cup\Omega_1}).$ 

\subsection{Uncorrelated case}

In the uncorrelated case the maps ${\cal R}^{(j)}$ define a (discrete time) dynamical system on $\mathbf{C}{\cal T}_{\Lambda}:$
the element $T$ is mapped at ``time"$=k\in\mathbf{N}$  into $T(k):={\cal R}^{(k)}\circ {\cal R}^{(k-1)}\cdots{\cal R}^{(1)}(T).$ In particular in the time independent
case $T(k)={\cal R}^k(T).$ In this section we will focus on the latter case. Since the eigenvalues of ${\cal R}$ are in $[0,1]$ (see v) in  Prop. 0)
the asymptotic behavior at $k=\infty$ is controlled by the eigenspace of ${\cal R}$ with eigenvalue $1$
i.e., ${\rm{Fix}}({\cal R}):=\{X\in{\cal A}(\Lambda)\,/\, {\cal R}(X)=X\}.$ Next proposition describes the structure of this subspace
and gives an explicit formula for the  limit purity.

%{\bf{ We plan to analyze the structure of these dynamical flows at the abstract level as well as for physically relevant instances.}}
%Preliminary concrete results in this program have been obtained in Refs \cite{RQC1,RQC2}. 
%We now move to consider them.
% \vskip 0.5truecm
%Let us consider the CP map ${\cal R}=\sum_{\Omega\in{\cal L}} q(\Omega) {\cal R}_\Omega$ the asymptotic behavior at $t=\infty$ is controlled by the eigenspace of ${\cal R}$ with eigenvalue $1$
%i.e., ${\rm{Fix}}({\cal R}):=\{X\in{\cal A}(\Lambda)\,/\, {\cal R}(X)=X\}.$
\vskip 0.5truecm
{\bf{Proposition 4}}%%%%%%%%%%%%%%%%%%%%%%%%%%%%%%%%%%%%% PROPOSITION 4

Let us consider ${\cal L}\subset 2^{\Lambda}$ as an {\em hypergraph} in $\Lambda$ and let $\{{C}_i\}_{i=1}^K$ be the family of its connected components. 
%${\cal L}=\bigcup_{i=1}^C {\cal L}_i$ 
\footnote{ A connected component in ${\cal L}$
is a maximal set $C_i\subset\cup_{\Omega\in{\cal L}} \Omega \subset\Lambda$ such that $\forall u, v\in C_i,\exists \Omega_1,\ldots, \Omega_n\in {\cal L}$
such that $u\in\Omega_1,\,v\in\Omega_n$ and $\Omega_{i}\cap\Omega_{i+1}\neq \emptyset, \,(i=1,\ldots,n-1)$
i.e., given any two vertices $u$ and $v$ there is a path of hyperlinks connecting them.
}
{\bf{i)}} 
\begin{equation}
{\rm{Fix}}({\cal R})=\bigcap_{\Omega\in{\cal L}} {\rm{Fix}}( {\cal R}_\Omega)=(\bigotimes_{i=1}^K {\mathbf{C}}\{ \openone, T_{C_i}\})\otimes {\cal A}((\cup_{i=1}^K C_i)^c)
\label{Fix}
\end{equation}
{\bf{ii)}} Using the notation of Eq. (\ref{pur-av}),  for a totally factorized initial state $\omega$
\begin{equation}
P_\infty:=\lim_{k\to\infty}\langle \omega^{\otimes\,2},\,{\cal R}^k(T_\Omega)\rangle=\prod_{i=1}^K\frac{ d^{|C_i| -|\Omega_i|}+d^{|\Omega_i|}   }{d^{|C_i|}+1 },\qquad(\Omega_i:=\Omega\cap C_i,\,i=1,\ldots, K)
\label{P_infSing}
\end{equation}

{\bf{Proof.--}} {\bf{i)}}  Let us first prove the first equality in (\ref{Fix}). If $X\in{\rm{Fix}}({\cal R})$ then 
$\|X\|=\|{\cal R}(X)\|=\|\sum_{\Omega\in{\cal L}} q(\Omega)  {\cal R}_\Omega(X)\|\le \sum_{\Omega\in{\cal L}}
q(\Omega)  \|{\cal R}_\Omega(X) \|\le
\sum_{\Omega\in{\cal L}} q(\Omega) \|{\cal R}_\Omega\| \|X\|=\left(\sum_{\Omega\in{\cal L}} q(\Omega) \right)\|X\|=\|X\|.$
Here we used $\|{\cal R}_\Omega\|=1,\,(\forall\Omega)$ and the normalization of $q$. 
This shows that all these inequalities are actually equalities that in turns implies that, $\|{\cal R}_\Omega(X)\|=\|X\|, (\forall\Omega),$
and, since the ${\cal R}_\Omega$'s are projections, ${\cal R}_\Omega(X)=X, (\forall\Omega\in{\cal L}).$ Therefore ${\rm{Fix}}({\cal R})\subset\bigcap_{\Omega\in{\cal L}} {\rm{Fix}}({\cal R}_\Omega).$
The reverse inclusion is obvious as $q$ is a  probability distribution. We now move to consider the second equality in (\ref{Fix}). We will here work at the level of the
algebra ${\cal A}(\Lambda)^{\otimes\,2},$ while in Appendix A the analysis is performed at the swap algebra ${\mathbf{C}}{\cal T}_\Omega$ level.

Let us begin by considering just two maps ${\cal R}_{\Omega_1}$ and ${\cal R}_{\Omega_2}.$ 
Since one has that Fix$({\cal R}_{\Omega_i})={\mathbf{C}}\{ \openone, T_{\Omega_i}\}\otimes {\cal A}(\Omega_i^c)^{\otimes\,2},\,(i=1,2),$
it is  easy to check  that  there are two cases: 
1) If $\Omega_1\cap \Omega_2=\emptyset$ then $F={\mathbf{C}}\{ \openone, T_{\Omega_1}\}\otimes{\mathbf{C}}\{ \openone, T_{\Omega_2}\}
\otimes A;$ and  2) If $\Omega_1\cap\Omega_2\neq \emptyset$ then $F={\mathbf{C}}\{ \openone, T_{\Omega_1\cup\Omega_2}\}
\otimes A$ where $F={\rm{Fix}}({\cal R}_{\Omega_1})\cap{\rm{Fix}}({\cal R}_{\Omega_2})$ and $A:={\cal A}((\Omega_1\cup\Omega_2)^c)^{\otimes\,2}.$
Case 1) is obvious, we will  then consider 2). 

Let $\Lambda_1,\,\Lambda_2$ and $\Lambda_3$ be  as in {\bf{ii)}} in the proof of Prop. 1 ($\cup_{i=1}^3\Lambda_i=\Omega_1\cup\Omega_2)$).
If $X\in{\rm{Fix}}\,{\cal R}_{\Omega_1}$ ($X\in{\rm{Fix}}\,{\cal R}_{\Omega_2}$) one can write $X=\alpha \openone_1\otimes\openone_2\otimes X_3+\beta T_1\otimes T_2\otimes Y_3$
($X=\alpha^\prime  X_1\otimes\openone_2\otimes \openone_3+\beta^\prime Y_1\otimes T_2\otimes T_3$) where $X_3$ and $Y_3$ ($X_1$ and $Y_1$) are arbitrary operators in ${\cal A}(\Lambda_3)^{\otimes\,2}$
( ${\cal A}(\Lambda_1)^{\otimes\,2}$). Therefore if $X\in F$ one must have
$\alpha \openone_1\otimes\openone_2\otimes X_3+\beta T_1\otimes T_2\otimes Y_3= \alpha^\prime  X_1\otimes\openone_2\otimes \openone_3+\beta^\prime Y_1\otimes T_2\otimes T_3,$
this equation can be solved for arbitrary $\alpha$ and $\beta$ iff
$\openone_1\otimes\openone_2\otimes X_3\sim X_1\otimes\openone_2\otimes \openone_3$ and $T_1\otimes T_2\otimes Y_3\sim Y_1\otimes T_2\otimes T_3.$
This implies $X_1=\openone_1$ and $X_3=\openone_3$ as well as $Y_1=T_1$ and $Y_3=T_3.$ In other terms the general element $X\in F$
must have the form $X=\alpha \openone_1\otimes\openone_2\otimes\openone_3+\beta T_1\otimes T_2 \otimes T_3$ namely $X\in {\mathbf{C}}\{ \openone_{\Omega_1\cup\Omega_2},\, T_{\Omega_1\cup\Omega_2}\}={\rm{Fix}}({\cal R}_{{\Omega_1\cup\Omega_2}}).$
This shows that $F\subset {\rm{Fix}}({\cal R}_{{\Omega_1\cup\Omega_2}}),$ the opposite inclusion is obvious, hence $F={\rm{Fix}}({\cal R}_{{\Omega_1\cup\Omega_2}}).$

The intersection over the whole ${\cal L}$ required in (\ref{Fix}) can be now perfomed, using 1) and 2), by organizing the  hypergraph $\cal L$  in to its connected components ${\cal C}_i, (i=1,\ldots,K).$
Different ${\cal C}_i$'s give rise, thanks to 1) to the different tensor factors in (\ref{Fix}); whereas intersections $\bigcap_{\Omega\in{\cal  C}_i} {\rm{Fix}}({\cal R}_\Omega)$
give rise, thanks to 2), to ${\rm{Fix}}({\cal R}_{C_i})={\mathbf{C}}\{\openone, \,T_{C_i}\},$ where $C_i:=\bigcup_{\Omega\in {\cal C}_i}\Omega.$
The last factor in (\ref{Fix}) simply reflects the fact that all the elements in the algebra over the set $(\bigcup_{i=1}^K C_i)^c$ are (trivial) fixed points for all the ${\cal R}_\Omega$'s.
(Notice that often this set will be empty). This completes the proof of (\ref{Fix}).
%By iterating these relations over all $\Omega\in{\cal L}$ one obtains te second equality in (\ref{Fix}). 

{\bf{ii)}}  Since $\cal R$ is a non-negative operator on the Hilbert-Schmidt space ${\cal A}(\Lambda)^{\otimes\,2}$ with unit norm its spectrum
Sp$({\cal R})=\{\rho_\alpha\}_\alpha$ is contained in $[0,\,1].$ Therefore ${\cal R}^k(T_\Omega)=\sum_\alpha \rho_\alpha^k F_\alpha \langle F_\alpha,\,T_\Omega\rangle,
\,( {\cal R}(F_\alpha)=\rho_\alpha F_\alpha)$ and $\lim_{k\to\infty} {\cal R}^k(T_\Omega)=\sum_{\alpha: \rho_\alpha=1}   F_\alpha \langle F_\alpha,\,T_\Omega\rangle.$
The relevant eigenoperators $F_\alpha$ can be chosen from an orthonormal basis of the space Fix$(\cal R)$ that we characterized in i). 
The latter eigenspace, according to i) has dimension at least $2^K$ but symmetry further simplifies the computation of $P_\infty.$ 
Notice indeed that, for a totally factorized state $\omega=\otimes_{i\in\Lambda}|\phi_i\rangle\langle\phi_i|,$ one has that
1) $\langle \omega^{\otimes\,2}, T_\Omega\rangle=1, (\forall\Omega\subset\Lambda)$ 
and 2) $\langle \omega^{\otimes\,2}, X_1 X_2\rangle=\langle\omega^{\otimes\,2}_1, X_1\rangle \,\langle\omega^{\otimes\,2}_2, X_2\rangle$
if $X_1$ and $X_2$ are supported on disjoint subsets $\Omega_1$ and $\Omega_2$ and $\omega_i=\otimes_{j\in\Omega_i} |\phi_j\rangle\langle\phi_j|, \,(i=1,2).$
These two facts imply in turn that, in each of the two-dimensional ${\mathbf{C}}\{\openone,\,T_{C_i}\}$ factors of ${\rm{Fix}}({\cal R}),$ only the symmetric term $\openone+T_{C_i}$ will have
give a non vanishing contribution to purity (as $\langle \omega^{\otimes\,2},\openone-T_{C_i}\rangle=0.$)
%i.e., after contraction with $\omega^{\otimes\,2}.$ 
Equation (\ref{P_infSing}) then follows by computing
the Hilbert-Schmidt scalar product of the swap  $T_\Omega$ with the normalized $\prod_{i=1}^C ( \openone+T_{C_i}). $
From  $\Omega=\left( \Omega\cap (\cup_{i=1}^K C_i)^c\right) \cup\left(    \Omega \cap (\cup_{i=1}^K C_i)  \right)=
\left( \Omega\cap (\cup_{i=1}^K C_i)^c\right) \cup \left( \cup_{i=1}^K \Omega_i\right)$
it follows that $T_\Omega =\left(\prod_{i=}^K T_{\Omega_i}\right) T_{\Omega\cap (\cup_{i=1}^K C_i)^c}.$
Since $\cal R$ has no non-trivial action on the last swap in $T_\Omega$
that term  (if present) gives rise to an irrelevant  factor one in $P_\infty.$ The other $K$ swaps instead give
$$P_\infty=\prod_{i=1}^K  \langle \frac{\openone+T_{C_i}}{\sqrt{2d^{|C_i|}(d^{|C_i|}+1)}},\, T_{\Omega_i}\rangle
\langle \omega^{\otimes\,2},\,\frac{\openone+ T_{C_i}}{\sqrt{2d^{|C_i|}(d^{|C_i|}+1)}}\rangle=
\prod_{i=1}^K  \langle \frac{\openone+T_{C_i}}{d^{|C_i|}(d^{|C_i|}+1)},\,T_{\Omega_i}\rangle,$$
 computing explicitly the scalar products above   
(use ${\rm{Tr}}(T_{\Omega_i})= d^{2(|C_i|-|\Omega_i|)}d^{|\Omega_i|},\, {\rm{Tr}}(T_{C_i} T_{\Omega_i})={\rm{Tr}}(T_{C_i-\Omega_i})= d^{|C_i|-|\Omega_i|} d^{2|\Omega_i|}$)
proves (\ref{P_infSing}). $\hfill\Box$
\vskip 0.5truecm

%This formula was first found in \cite{RQC1} on the basis of a conjecture. Prop.4 generalizes and rigorously proves that conjecture.
%\vskip 0.5truecm
{\bf{Remark.--}} As noticed in the above for a totally factorized state $\omega=\otimes_{i\in\Lambda} |\phi_i\rangle\langle\phi_i|$ one has that $\langle\omega^{\otimes\,2},\,T_\Omega\rangle=1,\,\forall\Omega$ regardless of the choice of the $|\phi_i\rangle$'s.
It follows  that in purity calculations $\omega^{\otimes\,2}$ can be replaced by the average
%\begin{eqnarray}
$$ \bigotimes_{i\in\Lambda}  \int dU_i U_i^{\otimes\,2}|\phi_i\rangle\langle\phi_i|^{\otimes\,2} (U_i^\dagger)^{\otimes\,2}=
\bigotimes_{i\in\Lambda}\frac{\openone+T_i}{d(d+1)}=\frac{1}{[d(d+1)]^{|\Lambda|}}\sum_{\alpha_1,\ldots,\alpha_{|\Lambda|}=0,1} T_1^{\alpha_1}\cdots T_{|\Lambda|}^{\alpha_{|\Lambda|}}=
%\nonumber \\
%\frac{2^{|\Lambda|}} {[d(d+1)]^{|\Lambda|}} \left( \frac{1} {2^{|\Lambda|}}\sum_{\Omega\subset\Lambda} T_\Omega\right)
:= \frac{1}{d_+^{|\Lambda|}}\,\Pi^+
$$
%\end{eqnarray}
 where $d_+:=d(d+1)/2$ and $\Pi^+:= 2^{-|\Lambda|}\sum_{\Omega\subset\Lambda} T_\Omega=
|{\cal T}_\Lambda|^{-1} \sum_{T\in{\cal T}_\Lambda} T .$ From this last expression is clear that $\Pi^+$ is the projector over the identity irrep of ${\cal T}_\Lambda$
acting on ${\cal A}(\Lambda)^{\otimes\,2}.$ Moreover, $\|\Pi^+\|^2_2={\rm{Tr}}\,\Pi^+=2^{-|\Lambda|}\sum_{\Omega\subset\Lambda} {\rm{Tr}}\,T_\Omega
=2^{-|\Lambda|}\sum_{\Omega\subset\Lambda} d^{|\Omega|} d^{2(|\Lambda|-|\Omega|)}=2^{-|\Lambda|} d^{2|\Lambda|}(1+d^{-1})^{|\Lambda|}=d_+^{|\Lambda|}\Rightarrow
\|\Pi^+\|_2=d_+^{|\Lambda|/2}.$
\vskip 0.5truecm
%%%%%%%%%%% EXAMPLE REM %%%%%%%%%%%%%%
{\bf{Example .--}} Let ${\cal G}=(\Lambda, \,E) $ be a connected graph and the ensemble map given by: ${\cal R}=|E|^{-1}\sum_{e\in E} {\cal R}_e.$ This is the so-called
random edge model considered in \cite{RQC1,RQC2}. Here we have that ${\cal L}$ coincides with $E$ (regarded as a family of subsets $\{v_1, v_2\}\subset\Lambda$)
therefore one has $K=1$ and $|C_1|=|\Lambda|$ and Eq (\ref{P_infSing})  becomes  $P_\infty=\frac{d^{|\Omega|}+ d^{|\Omega^c|}}{d^{|\Lambda|}+1}.$

%%%%%%%%%%%%%%%%%%%%% AREA LAWS %%%%%%%%%%%%%%%%%%%%%%%%%%
%{\bf{ Area laws.--}} 
\subsection{Area Laws for the uncorrelated case}
In this section we exploit the locality structure of the ensemble maps and the swap algebra relation (\ref{SA-inv}) to prove short-time
upper bounds for the average purity of a local region $A.$For simplicity, we will consider  a uniform distribution $q$ over $\cal L.$

From Eq. (\ref{SA-inv}) it follows that the maps ${\cal R}$ act on  $\mathbf{C}{\cal T}_{\Lambda}$ according the rule
${\cal R}(T_A)=\sum_{B\subset\Lambda} R_{B,A} T_B,\,(A\subset \Lambda)$ where 
\begin{equation}
R_{B, A:}=\sum_{\Omega\in {\cal L}\cap(\partial A)^c} q(\Omega)\, \delta_{B, A}+ \sum_{\Omega\in \partial A} q(\Omega)
\left(\alpha_+(A,\Omega)\, \delta_{B, A-\Omega} + \alpha_-(A,\Omega)\,\delta_{B, A\cup\Omega}\right).
\label{R-matrix}
\end {equation}
For a totally factorized initial state, one has  $P_k=\langle\omega^{\otimes\,2},\,{\cal R}^k(T_A)\rangle=
\sum_{B_1,\ldots, B_k} R_{B_k,B_{k-1}} R_{B_{k-1},B_{k-2}}\cdots R_{B_1, A}.$
%For $k=1$ Eq. (\ref{R-matrix}) gives
%$P_1=\sum_{B\subset\Lambda} T_{B,A}=\left( \sum_{\Omega\in {\cal L}\cap(\partial A)^c} q(\Omega)\right) +\sum_{\Omega\in\partial A}
%q(\Omega)\left(\alpha_+(A,\Omega)+ \alpha_-(A,\Omega)\right).$
 If $C:=|\Omega\cap A|=|\Omega\setminus A|$ then $c^-=0\Rightarrow \alpha_+=\alpha_-=c^+/2=d^C/(d^{2C}+1).$ 
%For uniformity with Refs \cite{RQC1, RQC2}  for $C=1$ the constant $c^+=2d/(d^2+1)$ will be denoted by $2N_d$
In the rest of the section we will assume that $\alpha_-=\alpha_+=:N_d,\,( \forall\Omega\in{\cal L},\,A\subset\Lambda)$
 \footnote{This is the case when the elements $\Omega_1$ in $\cal L$ are edges of a graph $\Omega_1=\{v_1,\,v_2\}\,(v_1,\,v_2\in\Lambda).$ In fact $\Omega_1\in\partial\Omega\Rightarrow \Omega_1\setminus\Omega=\{v_1\},$ and $\Omega_1\cap\Omega=\{v_2\}.$
Here  $N_d=d/(d^2+1)$\cite{RQC1}.}.  
The matrix (\ref{R-matrix}) can be written as a sum of a diagonal and a off-diagonal matrix $R=F_0+F_1$
where $F_0:={\rm{diag}}\,(1-p(A))_{A\subset\Lambda},\, ( p(A):=\sum_{\Omega\in \partial A} q(\Omega)=
1-\sum_{\Omega\in{\cal L}\cap (\partial A)^c} q(\Omega))$ and  $(F_1)_{B, A}=N_d \sum_{\Omega\in\partial A} q(\Omega)
\left( \delta_{B, A-\Omega} + \delta_{B, A\cup\Omega}\right).$ The off-diagonal matrix $F_1$ connects, with strength $N_d,$ each region $A$ with other 
$2|\partial A|.$
For example, for $k=1$ one immediately finds $$P_1=\sum_{B\subset\Lambda}  R_{B,A}=\sum_{B\subset\Lambda} ((F_0)_{B,A}+(F_1)_{B,A})=1-p(A) +2N_d\sum_{\Omega\in\partial A}q(\Omega)
=1-p(A)(1-2N_d),$$
For a uniform distribution $q$ over $\cal L$ one has $p(A)=|\partial A|/|{\cal L}|.$
In general: $P_k=\sum_{B\subset\Lambda}\left(  \sum_{\alpha_1,\ldots,\alpha_k=0}^1 F_{\alpha_1}\cdots F_{\alpha_k} \right)_{B,A}
%\sum_{[\alpha]\in{\mathbf{Z}}_2^k} \left(  \sum_{B\subset\Lambda} (F_{[\alpha]})_{B,A}    \right)
=:
\sum_{[\alpha]\in{\mathbf{Z}}_2^k} Y^{(k)}_A(\alpha).$
%For a uniform distribution $q$ over $\cal L$ one has $p(A)=|\partial A|/|{\cal L}|$ and, 
Using the matrix structure (\ref{R-matrix}), one obtains 
%\begin{equation}
$Y^{(k)}_A(\alpha)=\sum_{B\subset\Lambda} (F_{\alpha_1}\cdots F_{\alpha_k})_{B,A}
\le (2N_d\, p(X)  )^{|\alpha|} (1-p(\tilde{X}))^{k-|\alpha|}$
%\end{equation}
where: $|\alpha|:=\sum_{i=1}^k\alpha_i,$  $X$ ($\tilde{X}$) is the region in the family obtained by $A$ with $k$ successive allowed operations \footnote{
By allowed we mean that at each stage the element $\Omega\in{\cal L}$ added or subtracted to $A^\prime$ has to belong to $\partial A^\prime$} 
given by either joining or subtracting elements of $\cal L,$ with the largest (smallest) $|\partial X|$ ($|\partial \tilde{X}|$)\footnote{ Notice that $X$
and $\tilde X$ may depend on both $A$ and $k.$} Whence
%\begin{equation}
\begin{eqnarray}
P_k&\le& \sum_{|\alpha|=0}^k C_{k,|\alpha|} (2N_d\, p(X)  )^{|\alpha|} (1-p(\tilde{X}))^{k-|\alpha|}=\left(   1-p(\tilde{X})+  2N_d\, p(X)   \right)^k=
(1-(1-2N_d)p(X) +p(X)-p(\tilde{X}))^k\nonumber \\
&=&(1-e_p p(X))^k(1+\frac{\Delta_{k,A}}{1-e_p p(X)})^k\le
\exp\left[-k (p(X)e_p-\frac{\Delta_{k,A}}{1-e_p})\right]
\label{pur-bound}
\end{eqnarray}
where $C_{k,|\alpha|}$ are binomial coefficients,  $\Delta_{k,A}=p(X)-p(\tilde{X})$ and 
\begin{equation}
e_p:=1-2N_d=\frac{(d-1)^2}{d^2+1}.
\label{e_p}
\end{equation}
This constant is  the Haar average of the ``entangling power" for unitaries 
 acting on $d\times d-$dimensional systems introduced in \cite{ep}.
If we assume that each step the boundary size of a region can change just by 
a quantity $O(1)=o(|\partial A|)$ i.e., the hyper-graph ${\cal L}$ has bounded degree and the boundary size of $A$ is large scaling quantity, it follows $\Delta_k=O(k/|{\cal L}|)$ whereas 
$p(X)=O(|\partial A|/|{\cal L}|).$
In words: inequality (\ref{pur-bound})  describes, for $k=O(1),$ an exponentially bounded decay of average purity with a rate that fufills an {\em area law}
(up to  corrections $O(1)$) and that is proportional to the average entangling power $e_p$ \cite{RQC1}.
% in turn the lower bound on average entanglement (2-Renyi) entropy at step $k$
%\begin{equation}
%S_k\ge \frac{k}{|{\cal L}|}( |\partial X| E_p -\frac{c_k}{1-E_p})
%\end{equation}
%The constant $e_p$ is associated to the Haar average of the ``entangling power" for unitaries 
 %acting on $d\times d-$dimensional systems introduced in \cite{ep}.
%\vskip 0.5truecm
%%%%%%%%%%% EXAMPLE 1D
\subsection{An exactly solvable model: the uniform path-graph}
In this section we wil discuss a one-dimensional uncorrelated case where the use of swap algebra formalism allows for an explicit exponential reduction of computational complexity \footnote{ 
In \cite{RQC2} it  was  analyzed the case of the complete graph with $L$ vertices. Also in this case one has an exponential reduction of complexity as 
 average purity dynamics can be restricted to the $(L+1)$- dimensional ${\cal S}_L$-symmetric subspace of    $\mathbf{C}{\cal T}_\Lambda\cong (\mathbf{C}^2)^{\otimes\,L}.$ }.
The ensemble map $\cal R$ can be explicitly diagonalized in the relevant invariant sub-space of $\mathbf{C}{\cal T}_\Lambda$ and bounds on the convergence rate $P_k\to P_\infty$ are easily established. 

Let ${\cal G}=(\Lambda, E)$ be a {\em path-graph} of length $L:$
$\Lambda:=\{ 1,\dots,\, L\},\, E=\{ \{1,2\},\,\{2,\,3\},\ldots,\{L-1,\,L\}\}.$ From Eq (\ref{SA-inv}) 
it follows that the $(L+1)$-dimensional subspace of $\mathbf{C}{\cal T}_\Lambda$ spanned by $\{T_\Omega\,/\, \Omega=\{1,\ldots, i\},\,i=1,\ldots L\}$
and $T_\emptyset=\openone$  is invariant under all the edge maps ${\cal R}_e\,(e\in E).$
%
%{\bf{b)}} Let ${\cal G}$ the {\em complete graph} with $L$ vertices. In \cite{RQC2} it has been shown that purity dynamics can be restricted to the $(L+1)$- dimensional ${\cal S}_L$-symmetric subspace of   
%$\mathbf{C}{\cal T}_\Lambda\cong (\mathbf{C}^2)^{\otimes\,L}.$ 
%In these two cases one has exponential dimensional reductions $O(d^{2L})\rightarrow O(L).$ 
%Let us consider again the one-dimensional case discussed in . 
%
We denote the basis of this $(L+1)$-dimensional
invariant subspace by $\{|0\rangle,\,|1\rangle,\,|2\rangle,\ldots, |L\rangle|\}.$ 
In the uniform case one has ${\cal R}=\frac{1}{L-1}\sum_{i=1}^{L-1} {\cal R}_{\{i,\i+1\}}$ and relations (\ref{R-matrix}) give: 
\begin{equation}
{\cal R}|0\rangle= |0\rangle,\,{\cal  R}|L\rangle=|L\rangle,\qquad {\cal R}|i\rangle= a |i\rangle +b(|i-1\rangle+|i+1\rangle),\,(i=1,\ldots,L-1)
\label{1d}
\end{equation}
where $a:=(L-2)/(L-1)=1 -(L-1)^{-1}$ and $b= N_d(L-1)^{-1}.$ 
Let us denote by $R$ the (non-hermitean) matrix representation of $\cal R$ in the basis $\{|i\rangle\}_{i=0}^L.$ Eq (\ref{1d}) implies that $R$ is a sum of two matrices i.e., $R=R_1+R_2$
where $R_1=|0\rangle\langle 0| + b|0\rangle\langle 1| + |L\rangle\langle L| + b|L\rangle\langle L-1|$ and $R_2= a\openone_{L-1}  + b A_{L-1}.$
Here $\openone_{L-1}$ denotes the identity matrix in the $(L-1)-$ dimensional space spanned by $\{|i\rangle\}_{i=1}^{L-1}$ and $A_{L-1}$ is the {\em adjacency} matrix
of the path graph of length $L-1$ whose nodes are labelled by the same basis vectors.
%a \sum_{i=1}^{L-1} |i\rangle\langle i| +b
The matrix $R$ is diagonalizable by a (non-unitary) transformation $R=P^{-1}DP$ where $D={\rm{diag}}\,(\lambda_\mu)_{\mu=0}^L$ and $P$ is a matrix
whose columns are normalized eigenvectors of $R,$  i.e., $P_{j,\mu}=\langle j|\Psi_\mu\rangle, \, R |\Psi_\mu\rangle=\lambda_\mu |\Psi\rangle_\mu, (j,\mu=0,\ldots,L)$ \footnote{
Explicty one finds: $|\Psi_0\rangle=|0\rangle,\, |\Psi_L\rangle= |L\rangle$ and $|\Psi_h\rangle={\cal N}_h\left( \sum_{j=1}^{L-1} c_h(j) |j\rangle-b/\Delta_h\,(c_h(1) |0\rangle +  c_h(L-1)|L\rangle)\right),$
where $c_h(j):=\sin(\pi h j/L)$ for $h=1,\ldots,L-1.$ The normalization constant ${\cal N}_h$ is given by ${\cal N}_h=\left(L/2+ (b/\Delta_h)^2(c^2_h(1)+c_h(L-1)^2\right)^{-1/2}=O(L^{-1/2}).$
%Notice that $\sum_{j=0}^L P_{j,h}=\sum_{j=0}^L \langle j|\Psi_h\rangle={\cal N}_h\left( \sum_{j=1}^{L-1} c_h(j)-(b/\Delta_h)\,(c_h(1)+c_h(L-1))\right)=O(L^{-1/2}),$
%because both terms in the parentheses are $O(1).$
}
The spectrum of $R$ is:
%\begin{equation}
%{\rm{Sp}}({\cal R})=
$\{\lambda_\mu\}_\mu:=\{1,\, a+2b\cos(\frac{\pi h}{L}),\, (h=1,\ldots,L-1),\,1\}.$ The $L-1$ eigenvalues different from one can be written as
$1-\Delta_h$ where $\Delta_h:=\frac{1}{L-1}(1-2N_d \cos(\frac{\pi h}{L}))\in(0,\,1).$ The {\em spectral gap} $\Delta:= 1-\max\{ \lambda_h\}_{h=1}^{L-1}$ is then given by 
\begin{equation}
\Delta=\min_h \Delta_h=\frac{1}{L-1}\left(1-2N_d \cos(\frac{\pi }{L})\right)\ge \frac{1}{L}(1-2N_d)=\frac{e_p}{L}.
\label{spec-gap}
\end{equation}
Here $e_p$ is the average entangling power defined in Eq. (\ref{e_p}). The lower bound (\ref{spec-gap}) shows that the spectral gap of the ensemble map
$\cal R$ of the unifom path graph  is ``large" in the sense that it does not vanish faster than $L^{-1}$ for large system size $L.$ This fact, in turn
implies a fast convergence rate $P_k\mapsto P_\infty$ for $k\to \infty.$

For general $k$ and initial state $|l\rangle$ one has $P_k=\sum_{j=0}^L ({ R}^k)_{j,l}=\sum_{j=0}^L ({ PDP^{-1}})^k_{j,l}=\sum_{j=0}^L (PD^kP^{-1})_{j,l}=
\sum_{\mu=0}^L \lambda_\mu^k (\sum_{j=0}^L P_{j,\mu}) (P^{-1})_{\mu,i} .$ Since $\lambda_0=\lambda_L=1$  and $|\lambda_\mu|<1$ for $\mu=1,\ldots,L-1,$ one obtains
$P_k-P_\infty=\sum_{h=1}^{L-1} (1-\Delta_h)^k (\sum_{j=0}^L P_{j,h}) (P^{-1})_{h,l}.$ A direct computation shows that
\begin{equation}
P_k-P_\infty= \frac{2}{L}\sum_{h \,{\rm{odd}} } (1-\Delta_h)^k \sin(\frac{\pi l h}{L})\left[ \frac{2 N_d   \sin(\frac{\pi h}{L}) }{ 2N_d \cos( \frac{\pi h}{L}) -1} 
+\cot (\frac{\pi h}{2L})  \right]
\label{jed-result}
\end{equation}
where the sum is over the odd numbers $h$ between $1$ and $L-1,$ 
and, consistently with  (\ref{P_infSing}), $P_\infty =(d^{L-l}+d^l)(d^L+1)^{-1}$ \cite{jed}.
From this equation and the spectral gap definition (\ref{spec-gap}) a straightwforward  argument shows that $|P_k- P_\infty|\le l e^{-k\Delta} C$
where $C=O(1)$ is a constant depending on $d.$ Then, using the spectral gap lower bound in (\ref{spec-gap}), one obtains the convergence rate estimate
\begin{equation}
k\ge \frac{L}{e_p}( \log (\frac{C}{\epsilon}) +\log l)\Rightarrow  |P_k- P_\infty|\le\epsilon.
\label{conv-bound-1d}
\end{equation}
%From which $|P_k-P_\infty|\le (1-\Delta)^k \sum_{h=1}^{L-1}|\sum_{j=0}^L P_{j,h}|| (P^{-1})_{h,i}| \le (1-\Delta)^k (L-1)\,\max_h  \{|\sum_{j=0}^L P_{j,h}| \,|(P^{-1})_{h,i}|\}=O( (1-\Delta)^k L^{1/2} \|%P^{-1}\|) =O(L^{1/2} \|P^{-1}\| e^{-\frac{ke_p}{L}}).$ %\footnote{
%[We used: definition of spectral gap $\Delta,$ the estimate 
%$|\sum_{j=0}^L P_{j,h}|=O(L^{-1/2})$  (see [28]), $|(P^{-1})_{i,h}\|\le \|P^{-1}\|,$ $1-x\le e^{-x}\,$ and  $\Delta\ge e_p/L.$] From this rough  bound one obtains 
%\begin{equation}
%k\gg \frac{L}{e_p}\left(\log(1/\epsilon)+ 1/2\,\log L+\log \|P^{-1}\|\right)\Rightarrow |P_k-P_\infty|\le \epsilon.
%\label{conv-bound-1d}
%\end{equation}
%{\bf{How does $\|P^{-1}\|$ scale with $L$?}}
This inequality shows that, for fixed accuracy $\epsilon,$ the circuit length $k$ scales {\em linearly} with the total system size \cite{znidaric}.
Interestingly (\ref{conv-bound-1d}) also shows that  the convergence rate is proportional to the the average entangling power  $e_p,$ an intuitive result given the physical meaning of this quantity \cite{ep}.

We would like to end this section by showing that in this one-dimensional case  the bound (\ref{pur-bound}) (with $\Delta_{k, A}=0$) is tight  for  for short times $k$.
To begin with, notice that if $k \le \min \{ l, L-l\}$ the action of the map ${\cal R}^k$ on $|l\rangle$ does not involve the boundary states $|0\rangle$ and
 $|L\rangle.$ As long as these latter fixed states can be discarded Eq. (\ref{1d}) shows that ${\cal R}$ acts as the sum of three commuting translation operators
i.e., ${\cal R}\cong \sum_{\alpha=0,\pm 1} c_\alpha T_\alpha$ where $T_\alpha|l\rangle=|l+\alpha\rangle, \,(l=1,\ldots,L-1; \alpha=0,\pm 1)$
and $c_0=a,\,c_{\pm 1}=b.$ Therefore ${\cal R}^k|l\rangle=\sum_{\alpha_1,\ldots\alpha_k=0,\pm 1} c_{\alpha_1}\cdots c_{\alpha_k} T_{\alpha_1+\cdots+ \alpha_k}|l\rangle.$ As remarked in the above, for a totally factorized initial state, each vector in the last formula gives a unit contribution to the average purity $P_k,$ hence $P_k=\sum_{\alpha_1,\ldots\alpha_k=0,\pm1} c_{\alpha_1}\cdots c_{\alpha_k}=(c_0+c_{-1} + c_1)^k=(a+2b)^k=(1-\frac{1}{L-1} e_p)^k.$
This simple result can be also checked by resorting to the explicit formula (\ref{jed-result}).
 
%\end{equation}

\subsection {Correlated case}
In this subsection we will consider an ensemble map of the form ${\cal R}=\prod_{\Omega\in{\cal L}} {\cal R}_\Omega.$
This is a natural generalization of the correlated family of 1-dimensional L-RQC  called the ``contiguous edge model" 
introduced and analyzed in \cite{RQC1,RQC2}.
In the contiguous edge model the graph vertex set is given by $\Lambda=\{-L,\,-L+1,\ldots, -1,0,1\ldots,L\}$ and the edge set $E$ as in Sect. C. %Example 1 {\bf{a)}}. 
 The L-RQC model is now defined as follows:
 At each discrete time $t=k$ one selects $2L$ (Haar) random $U_e$ and $\mathbf{U}$ is built according to a permutation  $\pi\in {\cal S}_{|E|},\,(|E|=2L)$ 
  ${\mathbf{U}}=U_{e_{\pi(1)} }U_{ e_{\pi(2)}}\cdots U_{e_{\pi(|E|)}}$ and  then the process is iterated at $t=k+1$ and so on.
Each $U_{e_j}$ is  unitary acting on the edge $e_j=(j,\,j+1)$ state space $h_j\otimes h_{j+1}\cong (\mathbf{C}^d)^{\otimes\,2}.$
% We refer to $\pi$ as the {\em schedule.}
 The ensemble map corresponding to the specified model is given by ${\cal R}_\pi:= {\cal R}_{e_{\pi(|E|)} }\circ\cdots {\cal R}_{e_{\pi(1)} }.$ 
Concerning the map fixed point, in \cite{RQC2} it was  conjectured that  for {\em all}   $\pi\in{\cal S}_{|E|}$ the relevant fixed point of ${\cal R}_\pi$  is given by $\openone+T_{\Lambda}$
and $P_\infty$ by (\ref{P_infSing}). In the following we will demonstrate a result (Prop. 5) that, as a particular case, proves these conjectures.

Notice that, since in general different ${\cal R}_\Omega$ do not commute, the map ${\cal R}=\prod_{\Omega\in{\cal L}} {\cal R}_\Omega$
it is not even necessarily Hermitean. Therefore one needs a different type of analysis respect to the uncorrelated case of Prop. 4.
Let us start by summarizing in a Lemma some well-known facts and their proof.
\vskip 0.5truecm
{\bf{Lemma}}

Let $P_1,\ldots, P_n$ be projectors in a  finite-dimensional Hilbert Space, $R=:P_n P_{n-1}\cdots P_1$ and $F:=\cap_{i=1}^n {\rm{Im}}\,P_i.$
Then
{\bf{i)}} Fix$(R)=F,$
{\bf{ii)}} $\|R\|=1\Leftrightarrow {\rm{Fix}}(R)\neq \{0\},$
{\bf{iii)}} $R=P+Q$ where $P$ is the projection on $F$ and $\|Q\|<1.$
%$Q=Q_n Q_{n-1}\cdots Q_1$ where the $Q_i$'s are projections
%s.t. $Q_iP=PQ_i=0,\,(i=1,\ldots,n).$

{\bf{Proof.--}}
{\bf{i)}} Obviously $F\subset {\rm{Fix}}(R),$ let us show the reverse inclusion.
Let $x_0\in {\rm{Fix}}(R)$ and $x_i=P_i x_{i-1},\,(i=1,\ldots,n).$
One has $x_0=R(x_0)=x_n=P_n(x_{n-1})$ and then
$\|x_0\|=\|P_n x_{n-1}\|\le \|P_n\|\|x_{n-1}\|\le \|P_n\|\|P_{n-1}\|\|x_{n-2}\|\le\cdots\le\prod_{i=1}^n\|P_i\|\|x_0\|\le\|x_0\|,$
this shows that the equalities $\|P_i x_{i-1}\|=\|x_{i-1}\|, \,(i=1,\ldots,n)$ hold.
These latter, since the $P_i$'s are projections, imply $ x_i:=P_i x_{i-1} =x_{i-1}  \,(i=1,\ldots,n)$. Therefore
$x_n=x_{n-1}=\ldots=x_1=x_0,$ namely $P_i x_0= x_0$ for all $i$ i.e., $x_0\in F.$

{\bf{ii)}} If $F\neq\{0\}$ obviously $\|R\|=1;$ let us show the reverse implication.
If $\|R\|=1$ from compactness of the unit sphere $\exists x^*,\,\|x^*\|=1$ s.t. $\|Rx^*\|=\|x^*\|.$
From this relation, following the same steps as in i), one sees that $P_i(x^*)=x^*$ for all $i=1,\ldots, n$
i.e., $0\neq x^*\in F.$ Since $\|R\|\le\prod_{i=}^n\|P_i\|=1,$ if $F=\{0\}$ then
$\|R\|<1.$

{\bf{iii)}} $F$ is a subspace of all the spaces Im$P_i, \,(i=1,\ldots,n)$ hence one can write $P_i=P+Q_i$
where $Q_i$ projects on the orthogonal complement (in Im$P_i$ )of $F.$ 
Now we have $R=(P+Q_n)(P+Q_{n-1})\cdots (P+Q_1)= P +Q_n Q_{n-1}\cdots Q_1=: P+Q$
where we have used $Q_iP=PQ_i=0,\,(i=1,\ldots,n).$  Notice now that $Q=R(1-P)$
therefore $Q$ is a product of projections that (by construction) cannot have a non-trivial joint
fixed eigenspace. From ii) it follows that $\exists\Delta\in(0,\,1)$ s.t. $\|Q\|=1-\Delta.$
$\hfill\Box$
\vskip 0.5truecm%%%%%%%%%%%%%%%%%%%%%%%%%%%%%%%%%%%%%%%%%%%%%%%%%%%%%%%%%%%%%%%%%%%% PROP 5 %%%%%%%%%%%%%%
{\bf{Proposition 5}}

{\bf{i)}} Let ${\cal R}={\cal R}_{\Omega_{|{\cal L}|}}{\cal R}_{\Omega_{|{\cal L}|-1}}\cdots {\cal R}_{\Omega_1}, \,(\{\Omega_i\}_{i=1}^{|{\cal L}|}:={\cal L}) $ be the  ensemble map
of a correlated L-RQC. For a pure  totally factorized initial state $\omega$ one has
\begin{equation}
P_k=\langle\omega^{\otimes\,2}, {\cal R}^k(T_\Omega)\rangle=P_\infty+O(2^{|\Lambda|/2}e^{-Ak}),\quad A:=\log\frac{1}{1-\Delta},
\label{P_infSing1}
\end{equation}
where $P_\infty$ is given by (\ref{P_infSing}) and $1-\Delta=\|{\cal R}(1-{\cal P})\|<1,$
${\cal P}$ projector on Fix$({\cal R})$ (see Prop. 4).

{\bf{ii)}}  If ${\cal R}_\pi:= {\cal R}_{\Omega_{\pi(|{\cal L}|)} } {\cal R}_{\Omega_{\pi(|{\cal L}|-1)}}\cdots {\cal R}_{\Omega_{\pi(1)}}$
has the same infinite time purity Eq. (\ref{P_infSing1}) $\forall \pi\in{\cal S}_{|{\cal L}|}.$

{\bf{Proof.--}}
{\bf{i)}} We are in the setting of the Lemma with $F\neq \{0\}$ (Prop. 4) where now the finite-dimensional Hilbert space is
the Hilbert-Schmidt space ${\cal L}(\Lambda)^{\otimes\,2}$ the projections are the ${\cal R}_\Omega$ and $n=|{\cal L}|.$

One has ${\cal R}^k=({\cal P}+{\cal Q})^k={\cal P}+{\cal Q}^k$
(${\cal P}{\cal Q}={\cal Q}{\cal P}=0$).  Since $\|{\cal Q}^k\|\le \|{\cal Q}\|^k= (1-\Delta)^k=\exp\left( -k \log\frac{1}{1-\Delta}\right)$
it follows that $\lim_{k\to\infty} {\cal Q}^k=0$ and hence $\lim_{k\to\infty} {\cal R}^k={\cal P}.$
Therefore
$P_k=\langle\omega^{\otimes\,2}, ({\cal P}+ {\cal Q}^k)(T_\Omega)\rangle= P_\infty + \langle\omega^{\otimes\,2}, {\cal Q}^k(T_\Omega)\rangle,$
where $P_\infty=\langle\omega^{\otimes\,2}, {\cal P}(T_\Omega)\rangle$ is given (see Prop. 0)  by Eq. (\ref{P_infSing}). 
Now, using the Remark after Prop. 4,  and  Schwarz inequality for the Hilbert-Schmidt scalar product  we can write $\langle\omega^{\otimes\,2}, {\cal Q}^k(T_\Omega)\rangle=d_+^{-|\Lambda|}\langle\Pi^+,\,
{\cal Q}^k(T_\Omega)\rangle\le d_+^{-|\Lambda|} \|\Pi^+\|_2 \|{\cal Q}^k(T_\Omega)\|_2\le d_+^{-|\Lambda|/2} \|{\cal Q}^k(T_\Omega)\|_2\le
d_+^{-|\Lambda|/2}\|{\cal Q}^k\|\|T_\Omega\|_2\le (\frac{2d}{d+1})^{|\Lambda|/2} \|{\cal Q}\|^k\le  (\frac{2d}{d+1})^{|\Lambda|/2} (1-\Delta)^k\le 2^{|\Lambda|/2} (1-\Delta)^k). $
Where we also used $\|T_\Omega\|_2^2= {\rm{Tr}}\, T_\Omega^2={\rm{Tr}}\,\openone_\Lambda=d^{2|\Lambda|}.$
%Now 
 %by using Schwarz inequality for the Hilbert-Schmidt scalar product one finds $ |P_k-P_\infty|\le \|{\cal Q}^k(T_\Omega)\|_2\le \|T_\Omega\|_2\|{\cal Q}^k\|\le d^{|%\Lambda|/2} \|{\cal Q}\|^k=
 %d^{|\Lambda|/2}  \exp\left( -k \log\frac{1}{1-\Delta}\right).$ (We used $\|T_\Omega\|_2^2= {\rm{Tr}}\, T_\Omega^2={\rm{Tr}}\,\openone_\Lambda=d^{|\Lambda|}.$)
This proves (\ref{P_infSing1}). 

{\bf{ii)}} Point i)  shows that infinite time purity $P_\infty$ depends just on the joint fixed space
$\cap_{i=1}^{|{\cal L}|} {\rm{Im}}\, {\cal R}_{\pi(i)}={\rm{Im}}\,{\cal P}.$ The latter is clearly independent on the permutation $\pi.$ $\hfill\Box$
\vskip 0.5truecm
In the decomposition ${\cal R}_\pi={\cal P}+{\cal Q}_\pi$ the second term {\em does} in general depend on $\pi$
Accordingly the convergence rate of $P_k$ to $P_\infty$ (related, by Prop. 6, to $\|{\cal Q}_\pi\|$) may depend on $\pi.$ 
This phenomenon has been numerically observed in \cite{RQC1,RQC2}.
From (\ref{P_infSing1}) it follows that
%\begin{equation}
%k\gg \frac{|\Lambda|/2\,\log 2 +\log(1/\epsilon)}{\log\frac{1}{1-\Delta}}\Rightarrow |P_k-P_\infty|=O(\epsilon).
%\label{conv_bound}
%\end{equation}
%From (\ref{conv_bound}) also follows
 $k\gg \Delta^{-1} \log(2^{|\Lambda|/2}/\epsilon)\Rightarrow 
|P_k-P_\infty|=O(\epsilon).$
This shows how the convergence rate $P_k\rightarrow P_\infty$ is controlled by the norm $\|{\cal Q}\|=1-\Delta.$
This norm is also the second largest singular value of $\cal R.$ The positive constant $\Delta$ is therefore just the difference between the two largest singular values
of the ensemble map $\cal R$
and it is often referred to as the spectral gap (see Sect C). 
From the practical point of view it is crucial to determine how the convergence varies with the ype of L-RQC and how the spectral gap scales
as function of $|\Lambda|$ \cite{znidaric,viola}.  In \cite{RQC1} and \cite{RQC2} are given numerical results for a few natural choices of $\pi.$
In particular for the so called expanding  $\bar\pi$ (see Sect V.A in \cite{RQC2}), arguably the less efficient at generating entanglement, it was  found $P_k= 2 (1-e_p)^k(1+e_p)^{-k}$
($e_p$ is defined in (\ref{e_p})
for $1\ll k\le |\Omega|$ and a cross over to a volume-law for $k=|\Omega|.$ 
%Here $e_p=1-2N_d$ is the  average ``entangling power " defined in \cite{ep} for unitaries acting on $d\times d-$dimensional system.
Moreover:  1) asymptotic purity is given by (\ref{P_infSing}) (for $K=1$)  and, 2) the spectral gap of ${\cal R}_{\bar\pi},$
 %(in the restricted   space described in Example  {\bf{a)}} after Prop. 3) 
was found (numerically) to have  a finite  limit, $1-(2N_d)^2,$ for $L\to\infty.$

We end this section by noticing that, for both the uncorrelated and correlated cases, Propositions 4 and 5 show that the fixed point problem for the ensemble maps ${\cal R}$
can be formulated as a ground state problem for the ensemble ``Hamiltonians" 
%\begin{equation}
${\cal K}:= \sum_{\Omega\in{\cal L}} ({\mathbf{1}}- {\cal R}_{\Omega}).$
%\label{Ham}
%\end{equation}
This are {\em local} non-negative operators acting on the Hilbert-Schmidt space ${\cal A}(\Lambda)^{\otimes\,2}\cong\otimes_{i\in\Lambda}
{\cal B}(h_i^{\otimes\,2}).$ In fact one has  ${\cal K}\Phi=0\Leftrightarrow {\cal R}_\Omega\Phi=\Phi, \forall \Omega\in{\cal L},$
namely the ground state manifold of ${\cal K}$ coincides with the $2^K$-dimensional space Fix$({\cal R}).$
It would then be interesting to see whether one can adapt to the present context the techniques developed in \cite{nacht,brandao}  to lower bound the spectral gap of local Hamiltonians as well as to see whether one can exploit in convergence rate estimates the so-called quantum detectability lemma \cite{lemma}.

\section{Asymptotic statistics and local state designs}
In this section we will establish a simple connection between the formalism and the results of this paper and the theory of quantum local $t-$designs \cite{viola, harrow,brandao}.

The asymptotic average purity value (\ref{P_infSing}) is $O(d^{-|\Omega^c|})$ away from the minimal possible one for the reduced state of the  region $\Omega$ i.e., $P_{\rm{min}}=d^{-|\Omega|}.$
This implies that for sufficiently large $k$ most of the states $\omega_\Omega({\mathbf{U}_k}):= {\rm{Tr}}_{\Omega^c} ({\mathbf{U}}_k\omega{\mathbf{U}}_k^\dagger)$  (${\mathbf{U}}_k$ RQC of length $k$) should be close  to the maximally mixed state $\openone_{\Omega}/d^{|\Omega|}.$ Indeed on can prove that if $k$ is such that $|P_k-P_\infty|\le\epsilon$ then \cite{RQC2}
\begin{equation}
\overline{ \| \omega_\Omega({\mathbf{U}_k}) - \frac{ \openone_{\Omega}}{d^{|\Omega|}}     \|_1    }^{{\mathbf{U}}_k}\le \sqrt{ d^{|\Omega|}}\sqrt{  \frac{ 1}{d^{|\Omega^c|}}+ \epsilon  }.
\label{av_dist}
\end{equation}
Therefore if $|\Omega^c|=(1+\alpha)|\Omega|\, (\alpha>0$) and $\epsilon= d^{-|\Omega^c|}$ one has that the average trace norm distance between states generated by the ensemble of RQCs
and the maximally mixed one is $O(d^{-\alpha|\Omega|/2}).$ 
%In view of the convergence bound (\ref{conv_bound}) for this suffices
%$k=O\left(\Delta^{-1}\log d(|\Lambda| +(1+\alpha)|\Omega|)\right) ,\,(\alpha>0).$
Using  Markov inequality, for large $|\Omega|,$ one sees that with high probability each state in the ensemble is {\em  locally nearly indistinguishable from the totally mixed one}.
Notice now that similar claims can be  made for the Haar measure as: $\overline{ {\rm{Tr}}\, \omega_\Omega(\psi)^2 }^\psi\le d_{|\Omega|}^{-1}+ d_{|\Omega|^c}^{-1}$ \cite{ep}
where   $\omega_\Omega(\psi):={\rm{Tr}}_{\Omega^c} |\psi\rangle\langle\psi|.$ 

All this suggests that statistics of {\em local}  observables in $\Omega$ over the L-RQC ensemble (for large $k$) should be a good approximation of the Haar's one. 
In order to see this one has to show  that the distance ${\cal D}_t:=\| \overline{ \omega_\Omega({\mathbf{U}_k})^{\otimes\,t}}^{{\mathbf{U}_k}} -\overline{ \omega_\Omega(\psi)^{\otimes\,t}}^\psi \|_1,
$ ($t$-positive integer) is small e.g., less than $\delta$ in some limit. If this is the case all local observables will have all the L-RQC statistical moments up to the $t$-th one
at most $\delta$ away from the corresponding Haar's one  \cite{harrow}. An ensemble of states having this property i.e., to provide a way to approximately sample the 
Haar distribution is called an $\delta$-approximate   {\em state $t$-design}. How this can be done by efficiently resorting just to local quantum circuits  is a problem currently at the centre of an intense attention in view of its applications to quantum information theory (see recent  \cite{brandao}). 
%We plan to clarify the relation  between the relevant L-RQC we identified in the early stage of the project and  $t$-designs: 

%{\bf{ What are the conditions that make a L-RQC family into an approximate $t$-design? How efficient is it?}}
\vskip 0.5truecm
{\bf{Proposition 6}} For $k=O\left(\Delta^{-1}\log d(|\Lambda| +(1+\alpha)|\Omega|)\right) ,\,(\alpha>0)$ the
L-RQC family is a  $\delta-$approximate state $t$-design for local observables  $A\in{\cal A}(\Omega)$ with $\delta=O(\sqrt{t e^{-\alpha|\Omega|}}).$ 

{\bf{Proof.--}}
For $k=O\left(\Delta^{-1}\log d(|\Lambda| +(1+\alpha)|\Omega|)\right)$  Eq. (\ref{P_infSing1}) implies $|P_k-P_\infty|\le d^{-(1+\alpha)|\Omega|}:=\epsilon$
Using triangular inequality, convexity of the trace norm, Eq. (\ref{av_dist})  and $\|\omega^{\otimes\,t} -\sigma^{\otimes\,t}\|_1\le \sqrt{t}\|\omega-\sigma\|_1$
(for all pairs of density matrices  $\rho$ and $\sigma$) one obtains
 ${\cal D}_t\le \| \overline{\omega_\Omega({\mathbf{U}_k})^{\otimes\,t}-  {\frac{\openone_{\Omega}}{d^{|\Omega|}}^{\otimes\,t}}}^{{\mathbf{U}_k}}\|_1
+\|   \overline{\frac{\openone_{\Omega}}{d^{|\Omega|}}^{\otimes\,t}- \omega_\Omega(\psi)^{\otimes\,t}}^\psi \|_1\le   
 \overline{ \| \omega_\Omega({\mathbf{U}_k})^{\otimes\,t}-  {\frac{\openone_{\Omega}}{d^{|\Omega|}}^{\otimes\,t}}\|_1}^{{\mathbf{U}_k}} 
 +  \overline{\|\frac{\openone_{\Omega}}{d^{|\Omega|}}^{\otimes\,t}- \omega_\Omega(\psi)^{\otimes\,t}\|_1}^\psi \le  \sqrt{t} \left( \sqrt{ \frac{d^{|\Omega|}}{d^{|\Omega^c|}}+d^{|\Omega|}\epsilon}+
\sqrt{\frac{d^{|\Omega|}}{d^{|\Omega^c|}}  }\right)=:\delta=O(\sqrt{t e^{-\alpha|\Omega|}} )$ $\hfill\Box$.
% Here we used $\|\omega^{\otimes\,t}- \sigma^{\otimes\,t}\|_1\le \sqrt{t} \|\omega- \sigma\|_1$ and the former bounds. 
%For any local observable $A\in{\cal A}(\Omega)$ the expectation  averaged with respect to the Haar distribution is $\mu^{\rm{Haar}}(A)={\rm{Tr}}(A)/d^{|\Omega|}$ then
%$$|\mu^{\rm{Haar}}(A)-\overline{{\rm{Tr}}_{\Omega} (A\omega_{{\mathbf{U}}_k})}^{{\mathbf{U}}_k} |\le \|A\|
%\overline{ \| \frac{ \openone_{\Omega}}{d^{|\Omega|}} -{\rm{Tr}}_{\Omega^c} (\omega_{{\mathbf{U}}_k})\|_1}^{{\mathbf{U}}_k}= O(\|A\|d^{-\alpha|\Omega|/2}).$$
%This shows that the L-RQC family, for  larger $k$ of some ``equilibration time", and $|\Omega|=(1+\alpha)^{-1} |\Omega^c|$  is a $t$-unitary $\delta$- approximate design %\cite{harrow} for local observables $A\in{\cal A}(\Omega)$. The efficiency question is, again, how the equilibration time scales with the size of the system, $t$ and $\delta.$

\section{Conclusions}
In this paper we provided a mathematical presentation of the approach to local random circuits discussed in \cite{RQC1, RQC2}.
We introduced different classes of local random quantum circuits and  showed that their  statistical properties are encoded into an associated family of completely positive maps.
Ensemble average of quantum expectation can be studied as a function of the circuit length and we proved infinite time results as well as convergence bounds. Remarkably, 
  average entanglement dynamics can be described by the action of ensemble  maps on operator algebras of permutations (swap algebras). Also in this case we proved infinite time results for the expectation value of the purity  of a local region  both for  uncorrelated and correlated local random quantum circuits and short time area-law bounds for the uncorrelated case. The swap algebra formalism is powerful as it allows in some cases an exponential reduction of the problem size i.e., $d^{2L}\mapsto L.$
We illustrated such a phenomenon by an exactly solvable one dimensional uncorrelated case.  
Finally we briefly discussed a connection with $t-$design \cite{harrow,brandao} for localized observables.

\begin{acknowledgements}
This research was supported by the ARO MURI grant W911NF-11-1-0268 and by NSF grant numbers PHY- 969969 and PHY-803304.
Useful input from I. Arad , A. Harrow, S. Santra and J. Kaniewski  is gratefully  acknowledged.  
 
\end{acknowledgements}

%\end{document}
\appendix
\section{Fixed swaps subalgebras}
In this section we will show how some of the facts used in the Proof. of Proposition. 4 can be established working entirely within the swap algebra formalism.
%Let us start by noticing that 
\vskip 0.5truecm

{\bf{Proposition A1}}

Le ${\cal R}_{\Omega_1}$ be  an ensemble map given by (\ref{SA-inv}), $\Omega_1,\Omega_2\subset\Lambda$ and ${\rm{Fix}}\,({\cal R}_{\Omega_1}):={\rm{span}}\,\{T_\Omega\,/\, {\cal R}_{\Omega_1}(T_\Omega)=T_\Omega\}.$
\begin{itemize}
\item[{\bf{a)}}] $\Lambda_1\subset \Lambda$ implies that ${\mathbf{C}}{\cal T}_{\Lambda_1}={\rm{span}}\{T_\Omega\,/\, \Omega\subset\Lambda_1\subset\Lambda\}$ is a subalgebra of ${\mathbf{C}}{\cal T}_{\Lambda}$
of dimension $2^{|\Lambda_1|}.$

\item[{\bf{b)}}] If $\Lambda_1\cap\Lambda_2=\emptyset\Rightarrow{\mathbf{C}}{\cal T}_{\Lambda_1\cup\Lambda_2}\cong{\mathbf{C}}{\cal T}_{\Lambda_1}\vee {\mathbf{C}}{\cal T}_{\Lambda_2}\cong {\mathbf{C}}{\cal T}_{\Lambda_1}\otimes {\mathbf{C}}{\cal T}_{\Lambda_2}.$

\item[{\bf{c)}}]  ${\rm{Fix}}\,({\cal R}_{\Omega_1})={\rm{span}}\,\{T_\Omega\,/\, \Omega\supset\Omega_1\vee \Omega\subset\Omega_1^c\}=
{\rm{span}}\,\{T_\Omega\,/\, \Omega_1\notin\partial\Omega\},{\rm{dim}}\, {\rm{Fix}}\,({\cal R}_{\Omega_1})= 2^{|\Lambda|-|\Omega_1|+1}$

\item[{\bf{d)}}] ${\rm{Fix}}\,({\cal R}_{\Omega_1})$ is a subalgebra of ${\mathbf{C}}{\cal T}_{\Lambda}$ isomorphic to ${\mathbf{C}}\{\openone,\,T_{\Omega_1}\}\otimes  {\mathbf{C}}{\cal T}_{\Omega_1^c}.$
%(dim${\rm{Fix}}\,({\cal R}_{\Omega_1})=2^{|\Lambda|-|\Omega_1|+1}$)

\item[{\bf{e)}}] $\Omega_1\cap\Omega_2=\emptyset\Rightarrow {\rm{Fix}}\,({\cal R}_{\Omega_1})\cap {\rm{Fix}}\,({\cal R}_{\Omega_2})\cong {\mathbf{C}}\{\openone,\,T_{\Omega_1}\}
\otimes{\mathbf{C}}\{\openone,\,T_{\Omega_2}\}\otimes {\mathbf{C}}{\cal T}_{(\Omega_1\cup\Omega_2)^c}$

\item[{\bf{f)}}] $\Omega_1\cap\Omega_2\neq\emptyset\Rightarrow {\rm{Fix}}\,({\cal R}_{\Omega_1})\cap {\rm{Fix}}\,({\cal R}_{\Omega_2})\cong{\rm{Fix}}\,({\cal R}_{\Omega_1\cup \Omega_2})
\cong {\mathbf{C}}\{\openone,\,T_{\Omega_1\cup\Omega_2}\}\otimes  {\mathbf{C}}{\cal T}_{(\Omega_1\cup\Omega_2)^c}.$

\end{itemize}

{\bf{Proof.--}}

{\bf{a)}} Obvious.

{\bf{b)}} Any element of $\Lambda_1\cup\Lambda_2$, when $\Lambda_1\cap\Lambda_2=\emptyset,$ can be written in unique way as $\Omega_1\cup\Omega_2$
where $\Omega_i\subset\Lambda_i\,(i=1,2).$ Therefore $T_{\Omega_1\cup\Omega_2}=T_{\Omega_1\Delta\Omega_2}=T_{\Omega_1} T_{\Omega_2}$
showing the first isomorphism. Moreover, the map $\iota(T_{\Omega_1\cup\Omega_2})= T_{\Omega_1}\otimes T_{\Omega_2}$ can be checked to provide the second one. [Notice also ${\rm{dim}}\,{\mathbf{C}}{\cal T}_{\Lambda_1\cup\Lambda_2}=2^{|\Lambda_1\cup\Lambda_2|}=
2^{|\Lambda_1|}\,2^{|\Lambda_2|}={\rm{dim}}\,{\mathbf{C}}{\cal T}_{\Lambda_1}\,{\rm{dim}}\,{\mathbf{C}}{\cal T}_{\Lambda_2}.$]
If $\Omega_i,\tilde{\Omega}_i\in\Lambda_i\,(i=1,2)$ one has $(\Omega_1\cup\Omega_2)\Delta (\tilde\Omega_1\cup\tilde\Omega_2)=
(\Omega_1\Delta\tilde{\Omega}_1)\cup (\Omega_2\Delta\tilde{\Omega}_2);$ it follows that $\iota(  T_{\Omega_1\cup\Omega_2}\,T_{\tilde\Omega_1\cup\tilde\Omega_2}  )=\iota(T_{(\Omega_1\cup\Omega_2)\Delta (\tilde\Omega_1\cup\tilde\Omega_2)})=
\iota(T_{(\Omega_1\Delta\tilde{\Omega}_1)\cup (\Omega_2\Delta\tilde{\Omega}_2)})= T_{\Omega_1\Delta\tilde{\Omega}_1}\otimes T_{\Omega_2\Delta\tilde{\Omega}_2}=(T_{\Omega_1}\otimes T_{\Omega_2})(T_{\tilde\Omega_1}\otimes T_{\tilde\Omega_2})=
\iota(T_{\Omega_1\cup\Omega_2} )\,\iota(T_{\tilde\Omega_1\cup\tilde\Omega_2} )$  i.e., $\iota$ is also an {\em algebra} isomorphism (not just a vector space one).

{\bf{c)}} Notice first that, in view of (\ref{SA-inv}), ${\rm{Fix}}\,({\cal R}_{\Omega_1})$ is a linear subspace of ${\mathbf{C}}{\cal T}_\Lambda$ that contains ${\rm{span}}\,\{T_\Omega\,/\, \Omega\supset\Omega_1\vee \Omega\subset\Omega_1^c\}$ (as for every element $\Omega$ one has $\Omega_1\notin\partial\Omega$). Moreover the latter subspace
has dimension $2^{|\Lambda|-|\Omega_1|+1}$ as every element of its basis has the form $\Omega^\prime$ or $\Omega^\prime\cup\Omega_1$ where $\Omega^\prime\subset \Omega_1^c.$ Since ${\cal R}_{\Omega_1}$ is a projector the dimension of ${\rm{Fix}}\,({\cal R}_{\Omega_1})$ i.e., the eigenspace of ${\cal R}_{\Omega_1}$ with eigenvalue one,
is given by ${\rm{Tr}}\, {\cal R}_{\Omega_1};$ using again (\ref{SA-inv}) one sees that this trace is in fact $2^{|\Lambda|-|\Omega_1|+1}$ (only the elements $\Omega$ such that 
$ \Omega_1\notin\partial\Omega$ contribute, each by a one). This proves the equalities in c).

{\bf{d)}} If $\Omega,\Omega^\prime$ are such that $T_\Omega,\,T_{\Omega^\prime}\in {\rm{Fix}}\,({\cal R}_{\Omega_1})$ there are four possibilities 1) $\Omega\subset \Omega_1^c\wedge \Omega^\prime\subset \Omega_1^c$ 2) $\Omega\supset \Omega_1\wedge \Omega^\prime\subset \Omega_1^c,$ 3) $\Omega\subset \Omega_1^c\wedge \Omega^\prime\supset \Omega_1,$ 4)
 $\Omega\supset \Omega_1\wedge \Omega^\prime\supset \Omega_1.$ In all these cases it can be directly checked that $\Omega\Delta\Omega^\prime$ either contains
$\Omega_1$ or is contained in $\Omega_1^c$ i.e., $\Omega_1\notin \partial(\Omega\Delta\Omega^\prime).$ Hence $T_\Omega\,T_{\Omega^\prime}= T_{\Omega\Delta\Omega^\prime}\in {\rm{Fix}}\,({\cal R}_{\Omega_1}).$ The isomorphism in d) follows from the fact (noticed in c)) that every element in ${\rm{Fix}}\,({\cal R}_{\Omega_1})$ can be written as $T_{\Omega_1\cup\Omega^\prime}= {\iota}^{-1}( T_{\Omega_1} \otimes T_{\Omega^\prime})$ or
$T_{\Omega^\prime}={\iota}^{-1}(\openone\otimes T_{\Omega^\prime}),\,(\Omega^\prime\subset\Omega_1^c).$

{\bf{e)}} If $\Omega\in  {\rm{Fix}}\,({\cal R}_{\Omega_1})\cap {\rm{Fix}}\,({\cal R}_{\Omega_2})$ (with $\Omega_1\cap\Omega_2=\emptyset$) there are four cases:
1) $\Omega\supset \Omega_1 \wedge \Omega\supset \Omega_2$, 2)  $\Omega\subset \Omega_1^c \wedge \Omega\supset \Omega_2$, 3)
$\Omega\supset \Omega_1\wedge \Omega\subset \Omega^c_2,$ 4) $\Omega\subset \Omega_1^c\wedge \Omega\subset \Omega^c_2.$
Is easy to see that these alternatives account for the four different types of terms one gets from the tensor product in {\bf{e)}} e.g,
$\iota^{-1}(T_{\Omega_1}\otimes T_{\Omega_2}\otimes T_{\Omega^\prime}) =T_{\Omega_1\cup\Omega_2\cup \Omega^\prime},\,(\Omega^\prime\subset(\Omega_1\cup\Omega_2)^c)$
corresponds to 1) and so on.

{\bf{f)}} If $\Omega_1\cap\Omega_2\neq\emptyset$ then cases 2) and 3) in the above are not allowed and one is left with  1) and 4); namely, $\Omega\supset\Omega_1\cup\Omega_2$
or $\Omega\subset \Omega_1^c\cap\Omega_2^c=(\Omega_1\cup\Omega_2)^c.$ This means $\Omega_1\cup\Omega_2\notin\partial\Omega$ that, thanks to {\bf{c)}},
is equivalent to say $T_\Omega\in {\rm{Fix}}({\cal R}_{\Omega_1\cup\Omega_2}).$ Last isomorphism follows from {\bf{d)}}. $\hfill\Box$\vskip 0.5truecm
%YA-GUGU!!!!
%%%%%%%%%%%%%%%% THE END %%%%%%%%%%%%%%%%%%%%%%%%%%%%%%%%%%%%%%%%%%%%%%%%%%%%%%%%%%%%%%%%%%%%%%%%%%
\end{document}